\documentclass[12pt,twoside]{article}

\usepackage{hyperref}
\usepackage{geometry}
\usepackage{xcolor}
\usepackage{fancyhdr}

\usepackage[labelfont=bf]{caption}
\usepackage{natbib}
\usepackage{graphicx}
\usepackage{float}
\usepackage{booktabs}
\usepackage{tabularx}
\usepackage{titling}
\usepackage{titlesec}
\titleformat{\section}{\normalfont\bfseries\filcenter}{}{0pt}{}
\titleformat{\subsection}{\normalfont\bfseries\filcenter}{}{0pt}{\itshape}
\titleformat{\subsubsection}{\normalfont\bfseries\filcenter}{}{0pt}{\itshape}
\titlespacing*{\section}
  {0pt}{-.1\baselineskip}{-.1\baselineskip}  
\titlespacing*{\subsection}
   {0pt}{-.1\baselineskip}{-.1\baselineskip}  

\usepackage{lipsum}
\usepackage[T1]{fontenc}
\usepackage[T1]{fontenc}
\usepackage[utf8]{inputenc}
\date{}
\providecommand{\keywords}[1]
{
   \small	
  \textit{\hspace{-1em} Keywords: } #1
}
\fancypagestyle{firstpage}{
  \fancyhf{}
  \fancyfoot[L]{\color{gray} \footnotesize Copyright \copyright 2025 (Author1, Author2, AuthorN). Licensed under the Creative Commons Attribution Non-commercial No Derivatives (by-nc-nd). Available at: \url{http://journalqd.org}}%
  \lhead{\color{gray} \footnotesize Journal of Quantitative Description: Digital Media 5(2025), 1–n}
\rhead{\color{gray} \footnotesize 2673-8813/2025.XXXXX}
}
\title{ \normalsize \textbf{Adolescence to Adulthood Online: Tracing Youth Digital Expression on Reddit Over 13 Years} \vspace{-1.5em}}
\author{\normalsize Tyler Chang\\ \normalsize Drexel University, USA 
\vspace{1em} \\ \normalsize Khushboo Patel \\ \normalsize Drexel University, USA 
\vspace{1em} \\ \normalsize Afsaneh Razi \\ \normalsize Drexel University, USA \vspace{1em} \\
\normalsize Shruti Phadke \\ \normalsize Drexel University, USA \\ \thanks{Author 1: tlc373@drexel.edu} \thanks{Author 2: kp3329@drexel.edu} \thanks{Author 3: afsaneh.razi@drexel.edu} \thanks{Author 4: sp3945@drexel.edu} \thanks{Date submitted: 2026-04-07}} 
\begin{document}
\maketitle
\thispagestyle{firstpage}
\vspace{-8em}
\begin{abstract}
  \noindent  Reddit is widely used in research on youth and social media, yet there has been limited systematic examination of the diversity of content produced by young users or how their participation changes over time. We present a descriptive quantitative analysis of 443,856 Reddit posts between 2010 and 2023 by authors who self-disclosed their age as 11–-24. Using topic modeling, longitudinal statistical analyses, and psycholinguistic measures, we identify nuances in the topics discussed by youth and examine how posting patterns vary across age groups over time. Our results document both stable themes and extreme shifts in youth discourse, offering a comprehensive benchmark for longitudinal characterization of youth participation on Reddit. We further discuss how our findings signal early social media adoption by youth for self-disclosure and the extent to which changes in youth discourse may mirror broader offline events and evolving youth concerns.
\end{abstract}

\keywords{\textit{Reddit, Longitudinal Analysis, Mixed Methods, Youth} \vspace{8ex}}

\section{Introduction}
\label{sec:new_intro}

Reddit has emerged as one of the more popular digital venues where youth engage both with one another and with the broader online world \cite{Fischer2025Building, Nagata2025prevalence}. Users aged 13–24 now represent the platform’s fastest--growing demographic \cite{Saini2025RedditStatistics}, accounting for approximately 35.1\% \cite{BusinessOfApps2025RedditStatistics} of the over 1 billion monthly users \cite{Saini2025RedditStatistics}. 
Unlike other social media platforms, where personal identities play a central role in users' experiences, Reddit champions the pseudonymous status of its users, a feature that often leads youth to speak more candidly about personally significant topics \cite{kahlow2024beyond}. When considered alongside robust evidence that youth vary the level at which they disclose sensitive information or socially unpopular opinions depending on which social media platform they are presently engaged with \cite{Stupinski2022, tang2024social, razi2020sext}, it follows that general studies of youth on social media cannot be simply transposed onto Reddit. This is further complicated by Reddit users' tendency to congregate into communities that span multiple subreddits \cite{Saini2025RedditStatistics}. Consequently, the topics with which youth engage over time cannot be defined by subreddit membership alone.
Thus, as Reddit grows in prominence, so too does the complexity of youths' shifting topical focus and the importance of developing a systematic understanding of how youths' engagement with the platform evolves over time.

Prior studies have investigated how youth engage with a variety of topics on Reddit, including epilepsy support \cite{medina2024ALWE}, sexual health \cite{pettyjohn2025m}, mental health \cite{simhadri2023csa, sit2024youth}, emotional distress \cite{dan2025exploring}, child abuse \cite{Shulman2023}, medical misinformation \cite{Pollack2022ObesityReddit}, and interpersonal toxicity \cite{Kumar2023Understanding}. Although valuable by themselves, these studies do not yet offer a comprehensive and empirical mapping of the broader topic landscape youth traverse on Reddit. While risks associated with the use of social media are well--established---including misinformation \cite{shelby2023sociotechnical}, harassment \cite{razi2023Sliding, Razi2022Understanding}, privacy violations \cite{thomas2021sok, shelby2023sociotechnical, chang2025opaque, jia2015risktaking}, sexual predation \cite{razi2023Sliding, rybnicek2013facebook}, increased stress \cite{shelby2023sociotechnical, Nova2021facebook}, social isolation \cite{shelby2023sociotechnical}, cyberbullying \cite{Alluhidan2024Teen, thomas2021sok}, and mental health triggers \cite{Alluhidan2024Teen}---understanding how these risks can manifest within the topics youth explore on Reddit requires longer--term descriptive analyses than presently exist in the research literature. Although previous longitudinal studies of Reddit have been conducted, these have either been limited to topics such as climate change \cite{janaswamy2024exploring}, medical topics \cite{johnson2022sexually,de2014mental}, and meme culture \cite{ford2023competition, Uthirapathy2023LDA_BERT_ClimateTwitter, valensise2021entropy}, or are not specific to youth \cite{Wu2025BERTopicImmunotherapy, Medvedev2018modelling, Saini2025RedditStatistics}. Critically, the topics with which youth engage on Reddit carry real--world implications. As shown by Zhang et al. \cite{zhang2021teens}, shifts in youth--authored Reddit posts are not only detectable but often reflect real--life disruptions, making a broad temporal understanding of both topic trajectories and compositions essential for interpreting youth activity on Reddit. Building on this, our work adopts an initially topic--agnostic approach and, in doing so, provides a temporally robust understanding of how the topics youth engage with on Reddit have evolved over 13 years.

We provide a quantitative, descriptive mapping of longitudinal youth topic trajectories, focusing on characterization rather than causal inference. Per the World Health Organization (WHO) classifications and given the major physical, cognitive, and social changes associated with different stages of youth \cite{WHO2014Adolescents}, we consider three main developmental stages of youth: \textit{pre-teens} (<13), \textit{teens} (13–17), and \textit{young adults} (18–24). This stratification captures the times during which the greatest shift in a youth's psychology occurs \cite{klein1990adolescence}, as identifying explicit similarities and differences between age groups is paramount to a realistic understanding of how youth operationalize Reddit discourses over time. Accordingly, this paper seeks to answer the following research questions:

\begin{description}
    \item [RQ1:] \textit{What topics have youth primarily discussed on Reddit over time, and how have the number and distribution of topics discussed by youth on Reddit changed over time?}
    \item[RQ2:] \textit{ (a) How do topic distributions differ across pre--teens (<13), teenagers (13--17), and young adults (18--24) on Reddit longitudinally? 
    (b) How salient are wellbeing and safety-related topics to each specific age group?}
    \item [RQ3:] \textit{ (a) How do the more prominent youth safety-related subtopics vary linguistically? (b) What substantial psychosocial patterns can be observed across these subtopics?}
\end{description}

We employed a regular expression--based inferential method for age and gender detection to estimate the demographic metrics for authors of youth--created posts, the benefits and challenges of which we address in the methods section. We utilized time series, descriptive statistics, large--scale topic modeling (BERTopic), qualitative content analysis, and Linguistic Inquiry and Word Count (LIWC) to examine both the content and temporal trends of youth-authored posts on Reddit. We demonstrate that the topics discussed by youth followed distinct trajectories, fluctuating in posting volume at different times and to varying degrees. We also show that youth age groups differ in the attention they give to particular topics, with those relating to sex, intimacy, and relationships showing the sharpest contrasts across age groups and over time. Finally, we provide an empirical description of the broad linguistic separability of youth safety--related subjects. In doing so, this work establishes a 13--year baseline of youth discourse on Reddit across developmental stages, providing a foundation for contextualizing future findings within broader temporal patterns of youth expression.

\section{Methods}
\label{sec:new_methods}

\subsection{Data Collection}
\label{sec:new_data_collection}

Given Reddit’s pseudonymity and the resulting candidness it affords users \cite{Knuth2018-jr}, the platform was well--suited for studying youths’ evolving discussions. Moreover, Reddit's long history of public data availability allows for the longitudinal analysis presented in the paper. Our data collection proceeded in two stages. We first compiled a list of youth--identifying, non--bot users who had explicitly disclosed their age and gender in at least one post they authored. Second, using this filtered user list, we collected all posts created by these users between 2010 and 2023, yielding the final dataset. Both stages are described in detail below.

\textbf{Identifying Youth Authored Submissions on Reddit:} The PushShift dataset \cite{Baumgartner2020Pushshift} contains 651,778,198 posts across 2,888,885 subreddits, spanning June 2005 to April 2023. We searched the entire Pushshift dataset using the expanded version of regular expression patterns for age and gender presented in previous work \cite{tigunova2020reddust}. For example, \citeauthor{tigunova2020reddust} use patterns such as \texttt{"I am <number> years old"} for age disclosure. We added to their patterns by incorporating colloquial identity expressions such as \texttt{"I am 21M from.."} or, \texttt{"My (18F) boyfriend (19M).."}. Posts were classified as youth--authored when authors explicitly self-identified as being 24 years old or younger, or indicated that they were in their early twenties, teens, or childhood. The age limit of 24 was selected based on three factors: the approximate timing of brain maturation \cite{nimh_teen_brain_2023}, the typical age of university completion \cite{oecd2023eag}, and contexts involving mandatory military service \cite{worldpop2025militaryservice}. Some posts included vague age references to past periods (e.g., "As a teen, I loved to cook"); these posts were excluded from the initial collection unless they also included an explicit indicator that the post's author was still a youth at the time of writing. We also applied regular expressions to identify posts with explicit gender disclosures (e.g., "I am a boy" or "As a daughter"). While Reddit is commonly used by non-binary, genderqueer or gender--fluid persons, the associated gender identity expressions are extremely diverse. This challenges the reliability of post detection.  Accordingly, we focused on two genders. We discuss the limitations of this approach at the end of the Discussion. Requiring users to disclose both age and gender was necessary to achieve the granularity required by our research questions. While we did not assume that age-- or gender--based patterns would necessarily emerge, this dual requirement allowed us to detect such patterns if they were present. This initial process resulted in a dataset of 135,765 posts. It should be noted that we do not claim to have covered \emph{all} youth--authored posts in our study. Due to the strong selection bias imposed by our inclusion criteria, we do not claim that the studied population is representative of all youth discourse on Reddit. Instead, we take the persistence of our identified patterns across 13 years and thousands of subreddits as evidence of the stability of the observed trends and strong internal consistency. Relatedly, while there is a growing trend of using predictive models to classify the age and gender of social media users \cite{cinus2025inference, oulahbib2026applying}, our regex and keyword--based approach aligns with prior scholarship on demographic detection methods used for Reddit datasets \cite{zhan2019underage, tigunova2020reddust, emmery2024sobr}. We favored the non--predictive model approach for two reasons: (1) it ensured that no hidden associations captured by a potentially biased predictive model violated our requirement that demographic disclosures be explicit, and (2) it allowed us to provide greater transparency in our inclusion--exclusion criteria.

\textbf{Removing bot Accounts:} Like all social media platforms, Reddit contains a significant number of bot accounts \cite{Mournet2023RedditBots, Varol2017SocialBots, Ferrara2016RiseSocialBots, Cresci2020DecadeBots}. As we aimed to understand how youth discuss topics online and the psychosocial features of such conversations, it was crucial that we remove as many bots, spam, and other illegitimate accounts prior to collecting the full dataset. Accordingly, we applied the following methodology: we first consulted bot-detection subreddits (\texttt{r/botwatch}, \texttt{r/BotList}, \texttt{r/webscraping}, \texttt{r/ModSupport}) and publicly available bot lists \cite{redditSuspiciousAccounts}. Using the bot account names disclosed in these sources, we applied regular expression--based filtering to the authors of posts in our collected dataset, removing posts authored by users whose usernames exactly matched or closely resembled names on these lists. This filtering included usernames containing terms such as "bot", "robot", "auto", and "mod". The last of these terms was included to exclude automated moderator accounts, all of which contain either “mod” or “moderator” in their usernames. Our bot detection method is akin to that of Automod, which employs regular expressions to identify objectionable content \cite{jhaver2019human}, and is widely used on Reddit. This process removed 223 accounts from our dataset.

\textbf{Additional Robustness Checks:} To improve the reliability of our dataset, we applied multiple filtering steps to remove likely automated, spam, or otherwise unreliable accounts. First, to identify potentially automated posting behavior, we removed accounts that authored 15 or more posts within a one--minute interval (281 accounts). In the absence of a verifiable external criterion, this threshold was chosen conservatively, corresponding to approximately four seconds per post, based on the minimal time required to initiate a post, compose a title and a body, and submit it. Second, we removed accounts with inconsistent self--reported ages across posts (145 accounts). For example, an account reporting ages of 15 in 2010 and 19 in 2012 was excluded. This step was intended to reduce the likelihood of including accounts providing unreliable age information, an important effort given the role age played in our analysis. We did not do a similar consistency assessment for gender, as users may have amended their stated gender over time without violating our inclusion criteria.
Third, we removed accounts with usernames that indicated other social media platforms (e.g., Telegram) or contained links to unknown domains (e.g., 10.432tlm.co), as well as accounts whose posts consisted solely of advertisements or spam (1,770 accounts). Usernames and unknown domains were detected using regular expressions, whereas spam or advertisement posts were identified using a combination of regular expressions and keyword searches for known product or sales platforms (e.g., eBay). Finally, to further exclude automated or duplicated content, we used cosine similarity to detect clusters of highly similar posts by text, removing accounts whose posts exhibited greater than 90\% similarity to other posts (2,554 accounts). This method aligns with approaches used to detect coordinated activity or copy--pasta accounts on online platforms \cite{phadke2024characterizing}. 

We also excluded posts from subreddits known to contain legally dubious content (e.g., \texttt{r/SexyForSale}) or that advertised sexual services \cite{Corradini2021-da, Cauteruccio2022extraction}. During manual review of the dataset, we found that many posts focusing on sexual services were spam or scam--like, which we assessed as being unreliable representations of youth discourse. Although we acknowledge that some of the youth discourse on Reddit may focus on sexual services, we nonetheless excluded such posts as the 
inclusion of such content could inadvertently expose users to legal risk, and to ensure that the dataset reflected general user behavior. 
Collectively, these filtering steps removed 4,973 accounts, ensuring that the remaining dataset minimized automated activity, spam, and unreliable self-reported ages while potentially preserving genuine user content for analysis. This resulted in a final list of 14,590 youth users.

\textbf{Collecting all Submissions by Youth Authors:} Using the list of 14,590 youth users, we filtered the PushShift dataset to collect all posts authored by these users, yielding 1,461,130 posts across 29,095 subreddits. 
As not all submissions authored by youth had age disclosures, we used the age disclosures identified in the earlier subsection to label all submissions with ``age at the time''. For example, if an author posted ``I am 15M..'' in 2014, their age in other posts one year later will be labeled as 16. Ages were rounded to whole numbers: if a post occurred six or more months after the detected age month, we rounded up; otherwise, we rounded down. Unless explicitly noted otherwise, we assumed that users’ genders were consistent across all posts. Of the 12,792 authors included in our final dataset, only 5 (0.04\%) indicated a change in gender. As such, while we acknowledge that gender identity can change, such cases were rare, aligning our dataset with prior findings \cite{Hansen2024-gq}. As before, we cross--checked calculated ages against any self--reported age disclosures, as prior research shows a substantial proportion of minors on social media admit to falsifying their age to gain access to sites or content \cite{Madden2013teens, Nagata2025prevalence}. Accordingly, when conflicts arose between reported and calculated ages ($n = 17,899$), we prioritized the calculated age. We then reapplied our prior bot detection methods to identify potential bot, spam, or otherwise unsuitable accounts. Given the greatly expanded dataset, reapplying the full data cleaning pipeline was essential to maintaining data integrity. Accordingly, posts containing only advertisements or links to other domains, as well as posts authored by suspected bot accounts, were removed, resulting in the exclusion of 1,017,274 posts from 1,798 unique accounts. Finally, to capture different stages of youth development, we grouped users into three categories: \textbf{pre-teens (<13), teens (13--17), and young adults (18+)}. Collectively, these steps ensured that our final dataset of 443,856 posts was both high--quality and reflective of genuine human--authored content. 

\subsection{Identifying Topics Discussed by Youth on Reddit}
\label{sec:topic_model_liwc}

BERTopic models have been widely used for topic modeling of social media content \cite{Wu2025BERTopicImmunotherapy, Janaswamy2024ClimateReddit, Uthirapathy2023LDA_BERT_ClimateTwitter, emnlp2024pitopic}, including studies focusing on perceptions of immunotherapy \cite{Wu2025BERTopicImmunotherapy} and generative AI \cite{zhang2023public}. Accordingly, we applied a BERTopic model to our dataset, which initially generated 194 topics with a coherence score of 0.67, indicating moderately high semantic consistency and making our model's performance on online data in line with prior research \cite{chakraborti2024we}. Coherence, as defined here, represents the average of coherence scores for each individual topic, aligning our methods with that of \cite{morstatter2018search, roder2015exploring}. Parameter configuration for the BERTopic model was established through a grid search optimizing for topic coherence and coverage. 
All posts were labeled with a topic number and a confidence score, reflecting the model's certainty in its label. To reduce redundancy and improve interpretability, we manually examined the top and bottom 15 posts for each topic. This approach was chosen to capture both prototypical (high--confidence) and atypical (low--confidence) exemplars, providing a balanced view of each topic's content. One author conducted the initial review, after which all authors reviewed the assessments to ensure consensus. Topics demonstrating semantic similarity were merged, resulting in 154 subtopics. To further consolidate the structure, two authors independently coded these subtopics into 24 high--level topics, reconciling disagreements through discussion. We did not predefine the number of high--level topics. Rather, the final number resulted from an iterative consolidation of lower--level topics, balancing interpretability with the avoidance of overlapping or overly broad groupings. This process ensured a structured, interpretable framework of topic groupings suitable for downstream analyses. The final topics are discussed in detail in the Results section (Section \ref{sec:topics_analysis}).

\subsubsection{Analyzing Linguistic Markers of Youth Topics on Reddit}
\label{sec:LIWC_method}
Understanding youth discourse on Reddit requires more than identifying discussion topics. The psycholinguistic properties of youth--authored posts inform how topics are framed and help to explain why certain subjects shift in popularity over time. As such, we examined the linguistic characteristics of the topics themselves. That said, not all topics are of equal import or impact to youth. From the 24 high--level topics, we selected seven focal topics---Crime, Law \& Sociopolitical Issues, Physical Health \& Medicine, Mental Health \& Well-Being, Sex \& Intimacy, Body Image, and Relationships \& Friendships/Family \& Parenting---for detailed analysis. These topics were prioritized because they were prevalent in the dataset and intuitively aligned with concerns relevant to youth on social media, including mental health \cite{sit2024youth, Alluhidan2024Teen}, safety \cite{razi2023Sliding, shelby2023sociotechnical, thomas_sok_2021}, and social development \cite{Nova2021facebook, rybnicek2013facebook}. These focal topics also align with more general literature on youth--related risks \cite{national2011science, mobley2013testing, richard1991risk, jessor2013problem}, where topics related to drug usage (including its association with criminal elements), sexuality, and family circumstances have been identified as key factors of risk.

We used Linguistic Inquiry and Word Count (LIWC) \cite{tausczik2010psychological, liwc_manual} to characterize the psycholinguistic features of youth discourse across the focal topics described above. LIWC is a widely used text analysis tool that maps language to psychologically meaningful categories such as power dynamics, emotional tone, moral framing, and social interaction. By applying LIWC to topic--specific discussions, we examined how psycholinguistic patterns vary across themes, enabling us to capture both the topics youth addressed and the manner in which they conduct these conversations. The use of LIWC to extract psycholinguistic features from Reddit posts is well--established, as researchers have previously used the tool to examine differing levels of emotional--linguistic homogeneity between teen and non--teen parents \cite{melanson2025linguistic}, identify linguistic distinctions across mental health conditions such as depression, anxiety, and schizophrenia \cite{Kim2023MentalHealthReddit}, patterns within safe versus unsafe youth Instagram conversations on sexual topics~\cite{razi2023Sliding}, and track emotional patterns in Reddit posts following job resignations during the Covid-19 pandemic \cite{ireland-etal-2023-sadness}. As such, LIWC was deemed sufficiently versatile to accommodate the broad scope of our topic modeling approach. For each post, the title and body were combined into a single text, as users were inconsistent in how they structured their content. We then applied the 2022 edition of LIWC to the complete dataset, extracting 117 features. 

\textbf{Statistical Analysis: }Most features, including LIWC scores and post word counts, were heavily right skewed, prompting the use of predominantly nonparametric statistics \cite{Politi2021Nonparametric}. Because the research questions addressed distinct forms of comparison across demographic groups, topics, and temporal patterns, multiple statistical procedures were employed, each aligned with the structure of the data being evaluated. All tests were conducted with a predetermined threshold of $\alpha = 0.05$ for defining statistical significance, and post-hoc corrections were applied where relevant.

To contextualize topic--level and longitudinal findings (RQ2), we first examined demographic variation. Differences in posting behavior between male and female authors were evaluated using Welch's t--tests and Mann--Whitney U tests. Welch's t--test was selected to compare group means while accommodating unequal variances and sample sizes, which were expected in observational social media data \cite{ben2022novel, olteanu2019social, manga2026handling}. Mann--Whitney U tests were used as a nonparametric alternative that does not assume normality \cite{wall2023mann}, allowing us to assess whether group differences were consistent across the full distribution rather than driven by mean differences alone. Age--group differences across pre--teens, teens, and young adults were assessed using Kruskal--Wallis tests \cite{mckight2010kruskal}, as posting behavior and linguistic features exhibited non--normal and potentially skewed distributions. When omnibus tests were significant, Dunn's post--hoc tests with Holm correction were used to identify pairwise differences while controlling the family--wise error rate \cite{holm1979simple, garcia2009study}. 

These analyses characterized developmental--stage variation and informed the interpretation of topic--level structure. To address longitudinal topic dynamics (RQ1), we examined variability in posting behavior across topics using the Brown--Forsythe test for equality of variance \cite{allam2026brown}. This test was chosen because it provides a robust assessment of heterogeneity in variability across groups using deviations from the median, making it less sensitive to non--normality and outliers than other variance tests. Where appropriate, post--hoc pairwise comparisons with Holm correction were conducted to identify topic pairs with significantly different levels of fluctuation.  

We further examined monthly variation in topic prevalence using Kruskal–Wallis tests. This approach was selected due to unequal sample sizes across time points and non--normal distributions of topic prevalence. Dunn’s tests were used for post--hoc comparisons to identify topics with distinct longitudinal trajectories. To examine linguistic and psychosocial variation across topic domains (RQ3), we used Kruskal--Wallis tests, followed by Dunn’s tests where appropriate, on LIWC features derived from each post’s combined title and body. LIWC variables are often zero--inflated, with a high proportion of posts containing no instances of a given category, resulting in skewed, non--normal distributions; accordingly, a rank--based, nonparametric approach was used. Given the large number of potential pairwise comparisons across topics and LIWC categories (n = 2,457), statistical significance alone was not considered sufficient for interpretation. We therefore incorporated an effect size threshold to identify substantively meaningful differences. Effect sizes were calculated using Cliff’s Delta \cite{meissel2024using}, a nonparametric measure appropriate for ordinal and non--normally distributed data. Following established guidelines, only comparisons with large effect sizes ($|\delta| > 0.47$) were retained for further analysis. This filtering resulted in 15 focal topic pairs differentiated across 6 LIWC categories, which are examined in detail in Section~\ref{sec:liwc_analysis}.


\subsection{Ethical Statement}
\label{sec:new_ethical_statement}
Our Institutional Review Board (IRB) made a determination that this study does not constitute human subjects research as the data is publicly available and there are no interactions with human subjects.
To ensure ethical handling of the data, we anonymized all posts and carefully removed or masked any potentially identifiable information before reporting.
\section{Results}
\label{sec:results}


\subsection{A Brief Overview of the Dataset}
\label{sec:overviewDataset}

The final dataset included 443,856 Reddit posts authored by 12,792 users across 27,768 subreddits. 

\begin{figure}[ht]
    \centering
    \includegraphics[width = 0.9\linewidth]{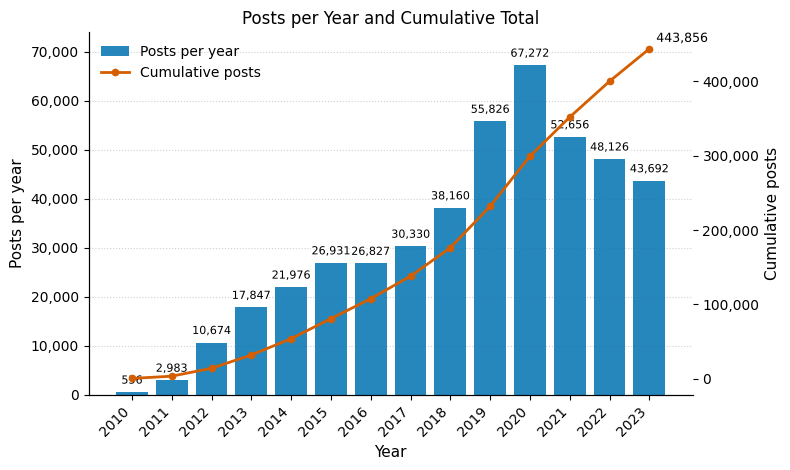}
    \caption{Posts per Year and Cumulative Total Diagram}
    \label{fig:postPerYEar}
\end{figure}

Of the 12,792 authors, 8,011 (62.6\%) self-identified as male and 4,781 (37.4\%) as female. The overall distribution of posts was 315,156 (71\%) authored by males and 128,700 (29\%) authored by females. While the average and median number of posts per user were approximately 36.7 and 7, respectively, these values differed significantly by gender: male users authored 39.34 posts on average compared to 26.89 for female users, with corresponding median values of 10 and 5. Welch’s \textit{t}-test (\textit{t} = 6.156, \textit{p} $<$ 0.001) and the Mann--Whitney \textit{U} test (\textit{U} = 21,586,186, \textit{p} $<$ 0.001) indicated that these differences were statistically significant\footnote{Due to the potential for significant lexical ambiguities, we have applied the following style guide to the Results and Discussion sections: (1) References to topics and subtopics that resulted from our use of BERTopic and manual coding are \textit{italicized}; (2) LIWC category names are written in \texttt{teletype ("typewriter") font}; (3) statistical variable names are designated as math objects and will appear as \textit{italicized} text.}.

\begin{figure}[ht]
    \centering
    \includegraphics[width = 0.9\linewidth]{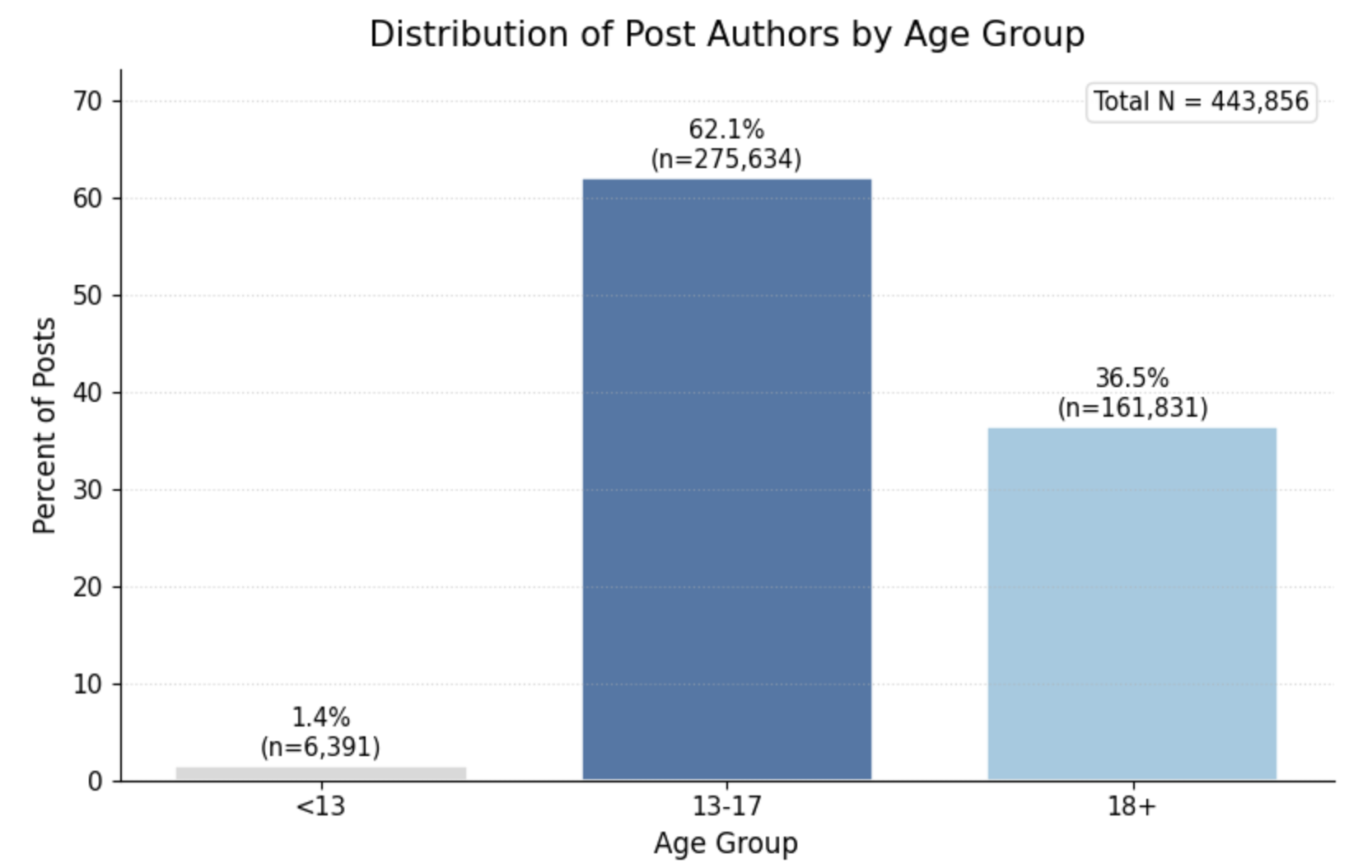}
    \caption{Distribution of the Age Groups}
    \label{fig:ageGroup}
\end{figure}

The ages of the users ranged from 11 to 24 years old, with posts made by 19--year--olds accounting for the largest group ($n$ = 131,110, 29.54\%). To capture different stages of youth development, we grouped users into three categories: \emph{\textit{pre-teens}} ($<$13), \emph{\textit{teens}} (13--17), and \emph{\textit{young adults}} (18+). The sizes of the groups varied substantially, with teens ($n$ = 275,634, 62.1\%) and young adults ($n$ = 161,831, 36.46\%) dwarfing pre-teens ($n$ = 6,391, 1.44\%). A Kruskal--Wallis test indicated significant differences in average posts--per--author distributions across age groups ($H$ = 10.398, \textit{p} = 0.0055). A subsequent Dunn’s post-hoc test with Holm correction (\textit{p} = 0.026) showed that the only significant difference was between teens ($mean$ = 33.84, $n$ = 8,091) and young adults ($mean$ = 35.97, $n$ = 4,549), suggesting broadly similar posting levels across youth groups with modest variation between older cohorts.

The posts spanned nearly fourteen years, with the first and last posts being created on January 2, 2010 and December 31, 2023, respectively. As shown in Figure \ref{fig:postPerYEar}, posting volume peaked in 2020 and generally declined thereafter. It should be noted that this distribution reflects youth--authored, nontrivial posts (see Section~\ref{sec:new_methods} for our definition of a trivial post), rather than overall posting activity on Reddit. According to Reddit's own transparency report for July--December 2023, over 4.4 billion pieces of content~\cite{reddit_transparency_2023b} were created during that six-month period alone, underscoring that the trends observed here reflect a specific subset of Reddit content and are not necessarily representative of overall platform--wide posting behavior.

\begin{figure}[h]
    \centering
    \includegraphics[width = 0.9\linewidth]{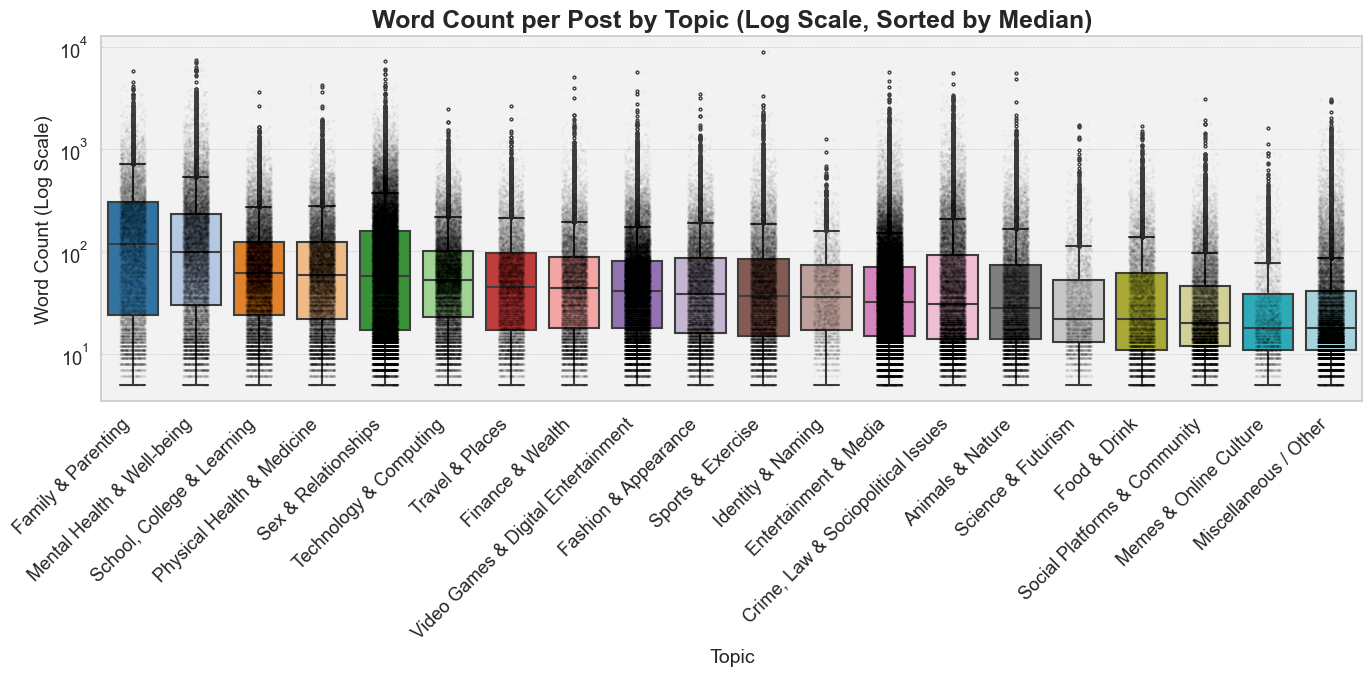}
    \caption{Word Count Distributions for each of the 24 Topics (log--scaled). The boxes are centered at the median word counts per topic and the block dots along the central vertical lines outside of the boxes' whiskers represent outliers. Horizontal dots indicate the density of the posts with a particular word count, with a darker cluster implying a higher number of posts.}
    \label{fig:wordCountDist}
\end{figure}

We also examined the word count distributions for each of our 24 topics. As shown in Figure \ref{fig:wordCountDist}, substantial outliers were present in each topic, indicating a highly skewed distribution. Together with the previously observed gender and age imbalances, this reinforced our use of medians and predominantly nonparametric statistical methods for subsequent analyses.

\subsection{Youth Topics of Discussion on Reddit}
\label{sec:topics_analysis}

The topic modeling methods described in Section \ref{sec:topic_model_liwc} resulted in 24 primary topics and 154 subtopics. A complete accounting of the topics, subtopics, keywords, and their respective prominence scores is available in the Appendix. The topics cover a wide variety of themes, ranging from video games and entertainment to mental health and relationships. 

\begin{figure}[ht]
    \centering
    \includegraphics[width = 0.8\linewidth]{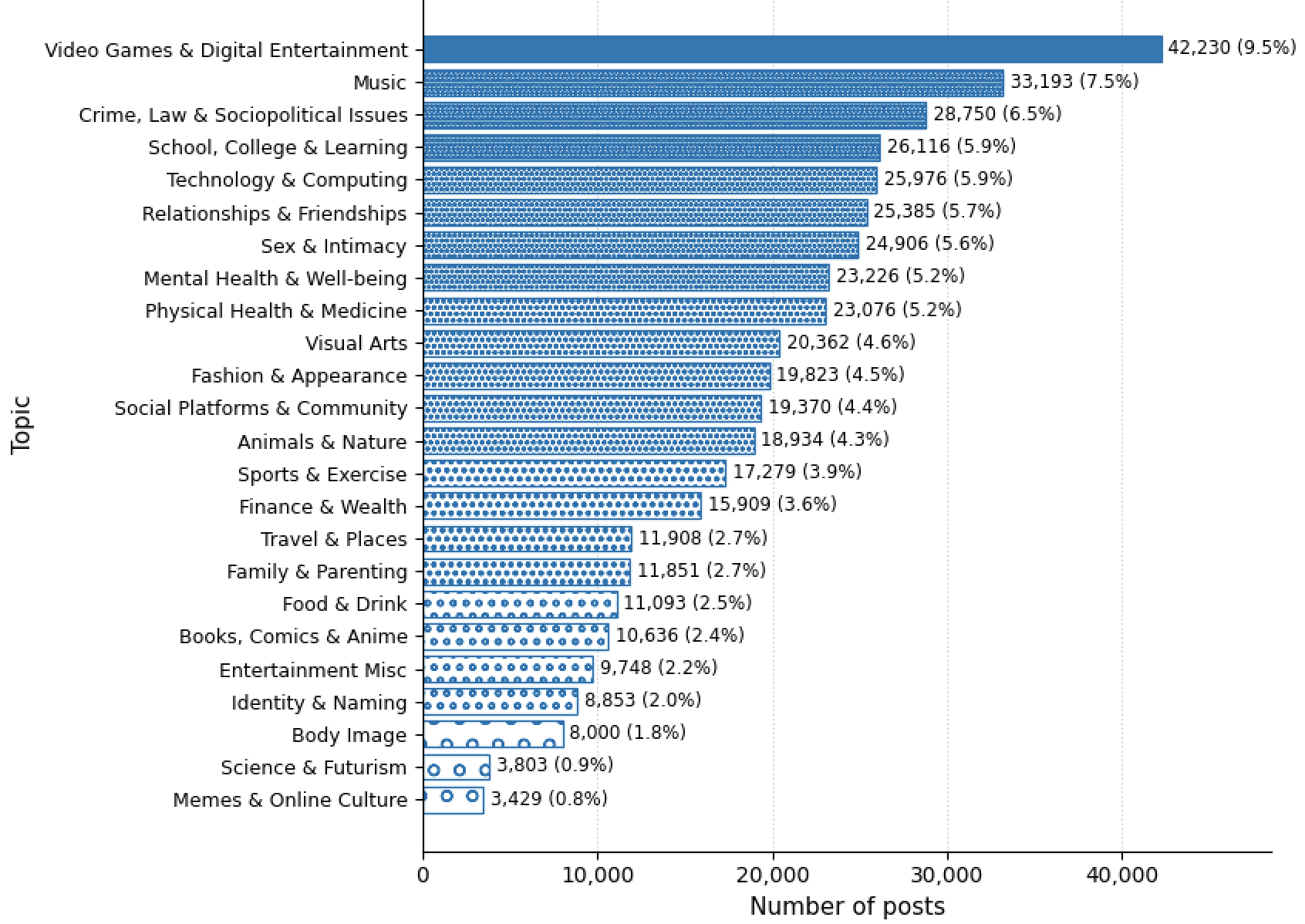}
    \caption{Number of Total Posts in each topic}
    \label{fig:distMajorTopics}
\end{figure}

\subsubsection{Temporal Variations in Post Volumes per Topic}
As shown in Figure \ref{fig:distMajorTopics}, the number of posts within each topic varied substantially, with topics having as few as 3,429 (0.8\%) and as many as 42,230 (9.5\%) posts. A Brown--Forsythe test comparing post counts across topics over time ($F = 15.27$, $p < 0.001$) indicated significant differences in variance, 
meaning that some topics exhibited greater fluctuation in posting volume over time than others. 
This variability refers strictly to changes in posting activity over time and does not imply differences in the internal composition of topics. A Kruskal--Wallis test on the absolute month--to--month changes post counts ($H = 402.79$, $p < 0.001$) further showed that the magnitude of increases and decreases in post volume varied significantly across topics. However, the frequency with which topics experienced increases or decreases did not differ significantly between topics (\textbf{Spikes:} $\chi^2$ \textit{p-value} = 0.469459; \textbf{Drops}: $\chi^2$ \textit{p-value} = 0.1937748). Simply put, while topics differed in how strongly the posting volume fluctuated, they did not differ in how often fluctuations occurred. 

Post--hoc analyses revealed that 121 of the 276 possible topic pairs (43.8\%) differed significantly in the variance of their monthly post counts according to the Brown--Forsythe test with Holm correction, and 102 of the 276 pairs (37.0\%) differed significantly in the distribution of their month--to--month changes in post counts according to Dunn's test with Holm correction following the Kruskal--Wallis test. These results indicated that a substantial share of topics differed both in the spread of their month--to--month variation and in the typical magnitude of those changes, suggesting heterogeneous temporal dynamics across topics. 

When restricting the analysis to the seven focal topics, the Brown--Forsythe post--hoc tests identified 8 of the 21 possible topic pairs (38.1\%) as differing significantly in their variance, indicating that certain focal topics exhibited greater fluctuation in month--to--month posting volume than others. The specific pairs showing significant variance differences are presented in Table~\ref{tab:bf_variances}.

\begin{table}[H]
\scriptsize
\centering
\caption{Significant Brown--Forsythe post--hoc comparisons of variance in monthly post counts across youth safety--related topics. Topic 1 exhibits higher variance than Topic 2. Variance values reflect the dispersion of monthly posting volumes over time. Reported p--values are Holm--corrected for multiple comparisons.}
\label{tab:bf_variances}
\begin{tabular}{l l r r r}
\toprule
\textbf{Topic 1 (Higher Variance)} & \textbf{Topic 2} & \textbf{Variance 1} & \textbf{Variance 2} & \textbf{P-Value} \\
\midrule
\textit{Crime, Law, \& Sociopolitical Issues} & \textit{Body Image} & 1967.85 & 194.11 & <0.000001\\
\textit{Relationships \& Friendships} & \textit{Body Image} & 1006.68 & 194.11 & 0.000045\\
\textit{Physical Health \& Medicine} & \textit{Body Image} & 1301.16 & 194.11 & 0.000083\\
\textit{Sex \& Intimacy} & \textit{Body Image} & 3711.42 & 194.11 & 0.000121\\
\textit{Mental Health \& Well-Being} & \textit{Body Image} & 758.15 & 194.11 & 0.000227\\
\textit{Crime, Law \& Sociopolitical Issues} & \textit{Family \& Parenting} & 1967.85 & 345.20 & 0.000001 \\
\textit{Sex \& Intimacy} & \textit{Family \& Parenting} & 3711.42 & 345.20 & 0.004397 \\
\textit{Physical Health \& Medicine} & \textit{Family \& Parenting} & 1301.16 & 345.20 & 0.042523 \\
\bottomrule
\end{tabular}
\end{table}

As shown in Table \ref{tab:bf_variances}, \textit{Body Image} exhibited significantly lower variability in its month--to--month posting volume when compared with five of the other focal topics, with \textit{Family \& Parenting} being the only exception. Relative to \textit{Crime, Law \& Sociopolitical Issues}, \textit{Relationships \& Friendships}, \textit{Physical Health \& Medicine}, and \textit{Mental Health \& Well-Being}, \textit{Body Image} showed between 3.91 and 10.14 times lower variance. Notably, \textit{Sex \& Intimacy} displayed substantially greater variability than \textit{Body Image}, with a variance approximately 19.13 times higher. These results suggest that \textit{Body Image} posting volume was comparatively stable over time, whereas several other focal topics exhibited greater temporal fluctuation in posting volume.

\textit{Family \& Parenting} also differed from several topics due to comparatively lower volatility in month--to--month posting volume. In comparison to \textit{Crime, Law \& Sociopolitical Issues} and \textit{Physical Health \& Medicine}, \textit{Family \& Parenting} showed between 3.77 and 5.7 times lower variance. This contrast was even more pronounced when compared with \textit{Sex \& Intimacy}, where variance was over 10.76 times lower. The absence of significant differences among the remaining focal topic pairs suggests that, for most youth safety--related topics, variability in posting volume followed broadly comparable patterns over time.

\begin{table}[H]
\scriptsize
\centering
\caption{Significant Dunn’s post--hoc comparisons of the distribution of absolute month--to--month changes in post counts across youth safety–related topics. Topic 1 exhibits a higher median change than Topic 2. Median values represent the typical magnitude of month--to--month variation in posting volume; p--values are Holm--corrected.}
\label{tab:kw_variances}
\begin{tabular}{l l r r r}
\toprule
\textbf{Topic 1 (Higher Median Change)} & \textbf{Topic 2} & \textbf{Median 1} & \textbf{Median 2} & \textbf{P-Value} \\
\midrule
\textit{Crime, Law \& Sociopolitical Issues} & \textit{Body Image} &  18 & 7 & <0.000001 \\ 
\textit{Mental Health \& Well-Being} & \textit{Body Image} & 13 & 7 & 0.019786\\ 
\textit{Physical Health \& Medicine} & \textit{Body Image} & 11 & 7 & 0.003545 \\ 
\textit{Relationships \& Friendships} & \textit{Body Image} & 11 & 7 &  0.006003 \\ 
\textit{Sex \& Intimacy} & \textit{Body Image} & 14 & 7 & 0.000121\\ 
\textit{Crime, Law \& Sociopolitical Issues} & \textit{Family \& Parenting} & 18 & 9 & 0.000001 \\
\bottomrule
\end{tabular}
\end{table}

A subsequent Dunn's test revealed that 6 of the possible 21 pairs (28.57\%) of youth safety--related topics differed significantly in the distribution of month--to--month posting changes. Consistent with the variance analysis, \textit{Body Image} and \textit{Family \& Parenting} emerged as comparatively distinctive topics. In particular, \textit{Crime, Law \& Sociopolitical Issues} typically exhibited a greater median magnitude of month--to--month change than \textit{Family \& Parenting}. Similarly, \textit{Body Image} showed a consistently lower median magnitude of change relative to the focal topics from which it differed significantly, with the only non--significant comparison being \textit{Family \& Parenting}. Together, these results signal that both the spread and the typical magnitude of posting fluctuations differed across focal topics, reinforcing the presence of heterogeneous temporal dynamics. 



\subsubsection{Temporal Variations in Post Volumes of Subtopics}
Recall that the 24 primary topics comprised 154 subtopics. To better understand which subtopics contributed to the most pronounced rises and falls within each focal topic (\textit{"Crime, Law \& Sociopolitical Issues," "Relationships \& Friendships," "Sex \& Intimacy," "Mental Health \& Well-being," "Physical Health \& Medicine," "Body Image"}, and \textit{"Family \& Parenting."}), we examined the distribution of subtopics over time. Table~\ref{tab:numSubperFocal} summarizes the number of subtopics within each focal topic, with the full list provided in Table~\ref{tab:topic_codebook_1}. In the following subsections, we describe instances of extreme spikes and drops in posting volume. A \textit{spike} or \textit{drop} was defined as three or more standard deviations from a topic's rolling 12--month average monthly post count. This threshold follows the three--sigma rule commonly used in time--series anomaly detection to identify statistically rare deviations from typical behavior \cite{braei2020anomaly}.

\begin{table}[H]
\scriptsize
\centering
\caption{Subtopics within each of the Focal Topics}
\label{tab:numSubperFocal}
\begin{tabular}{p{2.2cm} p{0.8cm} p{10cm}}
\toprule
\textbf{Topic} & \textbf{\#} & \textbf{Subtopics} \\
\midrule
\textit{Crime, Law \& Sociopolitical Issues} & 15 &
Africa; Bans; China \& Korea; Crimes \& Sentencing; Eurocentric Politics \& History; Islam; Law; Philosophy; Racism; Religion; Russian, German, and Southeast Asian Conflicts; Sociopolitical Topics; US Politics; Violent Conflicts; Weaponry \\
\midrule
\textit{Physical Health \& Medicine} & 10 &
Cutting \& Injuries; Dentistry; Digestive \& Bowel Issues; Drug Usage; Ear Pain; Eyes \& Vision; Injuries and Medical Advice; Organ Health; Sleep; Vaccines \& Covid \\
\midrule
\textit{Mental Health \& Well-Being} & 9 &
ADHD \& Neurological Conditions; Anxiety; Autism \& Neurological Conditions; Depression \& Poor Mental Health; Dreams; Fear; Mental Illness Medication; Motivation; OCD \\
\midrule
\textit{Sex \& Intimacy} & 8 &
Fantasy Roleplay; Nudes and NSFW Content; Periods \& Pregnancy; Physical Attractiveness; Porn \& Sexual Acts; Rape; Sexual Kinks \& Roleplay \\
\midrule
\textit{Body Image} & 3 &
Bodies; Body Height; Weight Loss \& Body Shaming \\
\midrule
\textit{Relationships \& Friendships} & 2 &
Friendship; Relationships \\
\midrule
\textit{Family \& Parenting} & 2 &
Children; Parent-Child Abuse \\
\bottomrule
\end{tabular}
\end{table}

\textbf{Crime, Law \& Sociopolitical Issues: }Comprised of 15 subtopics, including religion, violent conflicts, and US politics, \textit{Crime, Law \& Sociopolitical Issues} exhibited three extreme spikes in 2012, 2018, and 2020. The first spike coincided with an increase in posts related to \textit{religion}, particularly \textit{Islam}, with these two subtopics collectively accounting for approximately 35.85\% of the \textit{Crime, Law \& Sociopolitical Issues} posts made in 2012. A similar pattern was observed in 2018, where \textit{religion} (20.77\%) and \textit{US politics} (18.03\%) together represented over a third of all posts. The third spike, occurring in 2020, again corresponded with a rise in \textit{religion}--related posts (22.14\%). Across these events, fluctuations in posting volume were associated with shifts in the relative prominence of specific subtopics rather than changes in the overall structure of the topic.

\textbf{Physical Health \& Medicine:}
Unlike \textit{Crime, Law \& Sociopolitical Issues}, which exhibited only extreme spikes, \textit{Physical Health \& Medicine} showed both extreme rises and falls in posting volume. The first two extreme drops occurred in 2011 and 2014. In 2011, \textit{Drug Usage} accounted for 58.97\% of posts within the topic, while the \textit{Eyes \& Vision} subtopic—representing approximately 14.29\% of posts in 2010—was absent. A similar decline occurred in 2014, during which the share of \textit{Drug Usage} posts decreased from 58.97\% to 28.57\%. In 2017, \textit{Drug Usage} regained prominence, rising to approximately 35\% of posts and coinciding with an extreme spike in overall posting volume. A subsequent spike in 2020 corresponded with a sharp increase in posts related to \textit{Vaccines \& Covid}, with this subtopic surging from 3.57\% to 37.01\% of posts within the topic. The fluctuations in posting volume, therefore, aligned with changes in the relative prominence of specific health-related subtopics.

\begin{figure}[H]
    \centering
    \includegraphics[width = \textwidth]{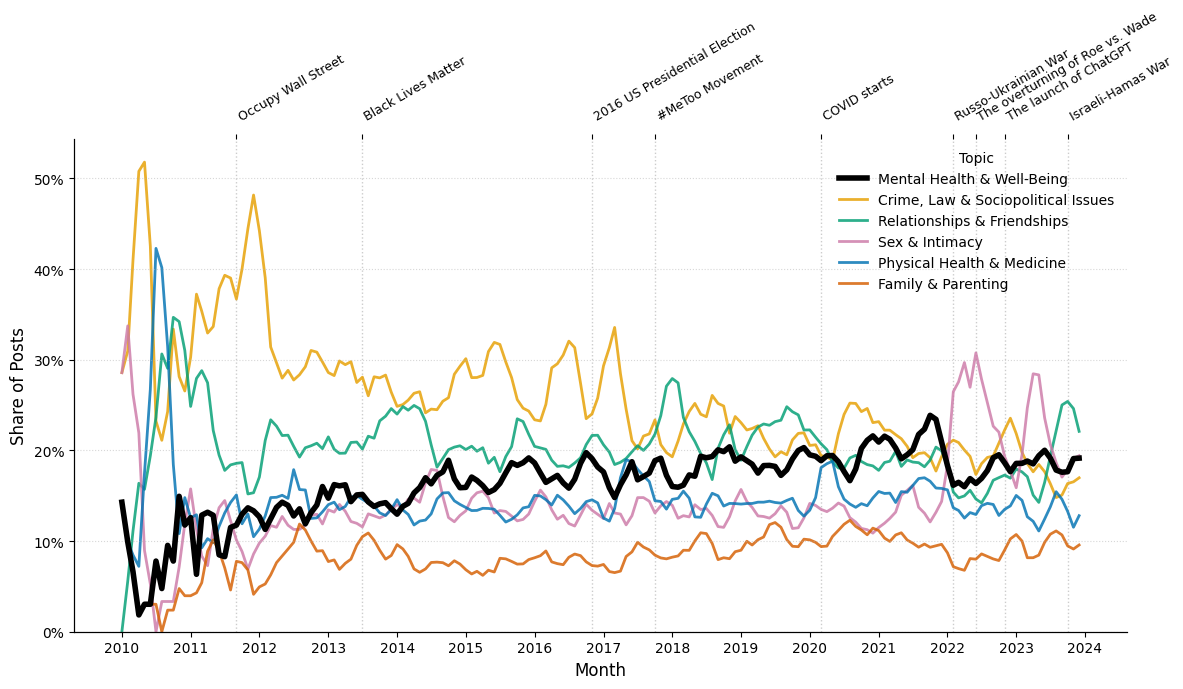}
    \caption{The black line indicates the mental health trend over time. Annotations show the dominant subtopic at the time. The other colored lines show the trends for the other topics over the same time period. Vertical dotted lines indicate the timing of major world events that were pertinent to the focal topics at the time. While most of the events are U.S.--centric, prior work has shown that approximately 35\% of Reddit engagement comes from U.S.–based users, making these events more likely to have significant impacts on Reddit content.}
    \label{fig:mentLinePlot}
\end{figure}

\textbf{Mental Health \& Well-Being:}
\textit{Mental Health \& Well-Being} exhibited five extreme events (Figure \ref{fig:mentLinePlot}), all of which were classified as drops. The first two occurred in 2012 and 2014 and coincided with the elevated prominence of the subtopic \textit{Fear}, which accounted for 27.5\% and 39.08\% of posts in those respective years. In 2012, the subtopic \textit{Dreams} showed similar prominence but declined in subsequent years. From 2016 through 2019 and 2020, \textit{Depression \& Poor Mental Health} was the dominant subtopic, accounting for between 41.32\% and 51.97\% of posts. Despite this increasing concentration in later years, the overall topic experienced multiple extreme drops across the study period, indicating that short-term declines in posting volume occurred even as the long--term prominence of the topic grew.

\textbf{Sex \& Intimacy:}
Two extreme spikes were observed in the \textit{Sex \& Intimacy} topic, with the first in 2012 and the second in 2022. The 2012 spike coincided with an increase in posts related to \textit{Porn \& Sexual Acts} which accounted for approximately 40\% of posts within the topic that year and broadly reflected overall posting trends among focal topics during the same period. The later spike corresponded with a substantial increase in posts addressing \textit{Sexual Kinks \& Roleplay}. While this subtopic represented 16.67\% of \textit{Sex \& Intimacy} posts in 2012, by 2022, it accounted for 69.66\%, indicating a marked shift in the internal composition of the topic over time.

\textbf{Body Image:}
The \textit{Body Image} topic exhibited an equal number of extreme spikes and drops, with two of each observed over the study period. Across all years, \textit{Weight Loss \& Body Shaming} remained the dominant subtopic, accounting for between 72.34\% (2016) and 84.62\% (2012) of posts within the topic. Notable drops occurred in 2011 and 2012, during which overall posting volume declined. In contrast, 2016 and 2017 showed increases in posting activity across all subtopics, coinciding with extreme spikes in those years. Despite these fluctuations, the relative prominence of the dominant subtopic remained consistently high throughout the study period.

\textbf{Relationships \& Friendships:}
The \textit{Relationships \& Friendships} topic exhibited five extreme events, four of which were spikes and one a drop. The drop, occurring in 2011, corresponded with a general reduction in posting volume rather than a shift in any particular subtopic. Across the study period, the non--platonic and romantic \textit{Relationships} subtopic consistently accounted for the majority of posts and coincided with the four subsequent spikes in 2012, 2013, 2017, and 2019. Although this subtopic remained dominant, its share of posts varied over time, ranging from 76.64\% (2013) to 88.67\% (2017), reflecting fluctuations in internal composition despite overall structural stability.

\textbf{Family \& Parenting}
The final focal topics, \textit{Family \& Parenting}, showed two extreme spikes and no extreme drops. The first spike occurred in 2012 and corresponded with a substantial increase in posts related to \textit{Parent-Child Abuse}, which accounted for 88\% of posts within the topic that year. The second spike did not occur until 2023, again coinciding with a high concentration \textit{Parent-Child Abuse} (87.28\%). It is important to note that \textit{Family \& Parenting} consisted of only two topics---\textit{Parent-Child Abuse} and \textit{Children}---with the former accounting for 90.56\% of posts overall, indicating a consistently concentrated composition.

\subsection{Age Groups Topics over Time}
\label{sec:time_and_ages_analysis}

\begin{figure}[H]
    \centering
    \includegraphics[width = 0.8\linewidth]{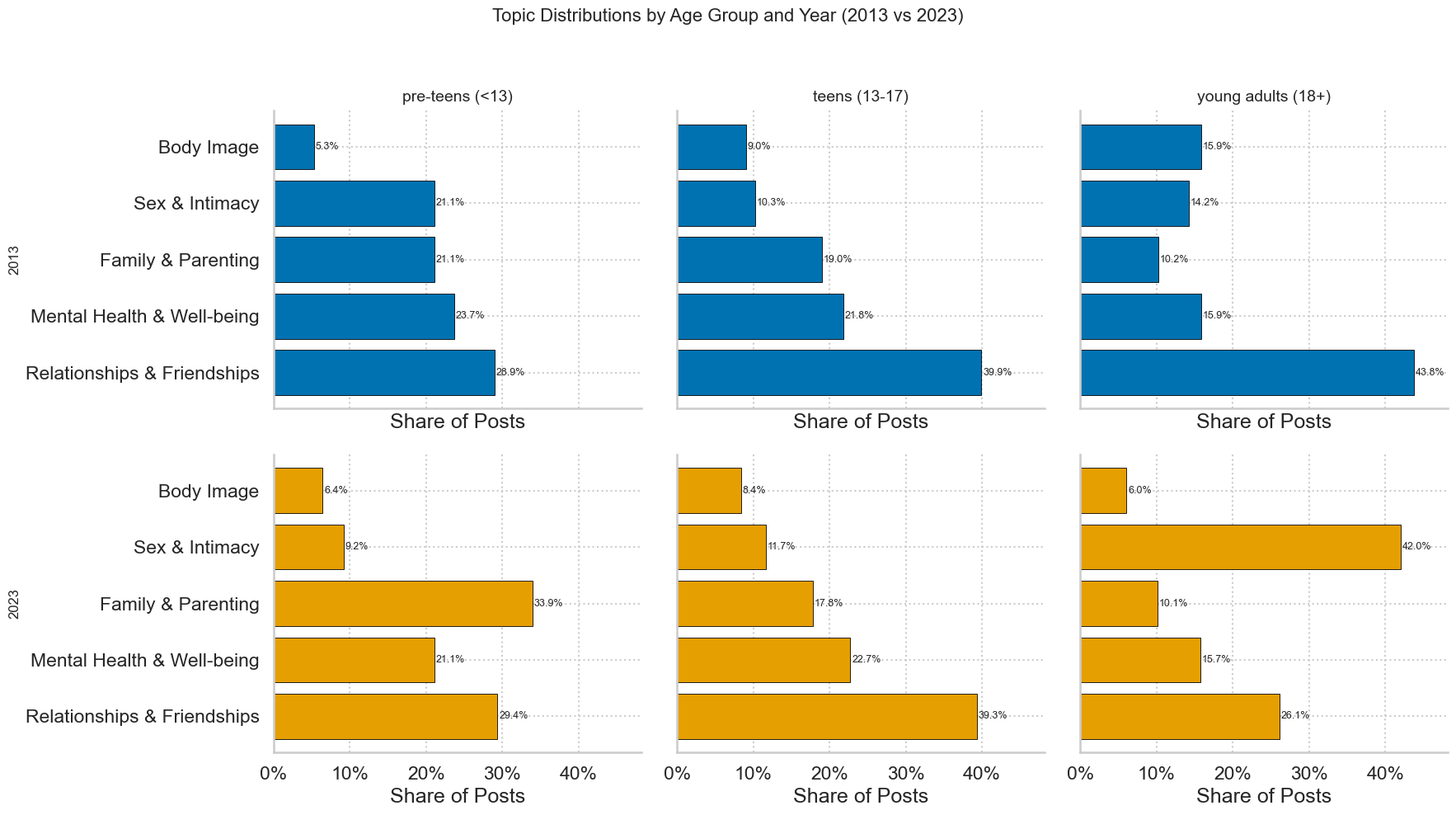}
    \caption{Focal Topic Distributions by Age Group and Year (2013 vs. 2023). Two of the focal topics (Crime, Law \& Sociopolitical Issues and Physical Health \& Medicine) did not receive any posts in 2013 or 2023.}
    \label{fig:2013v2023ageYear}
\end{figure}

While the last section describes temporal trends in topics and subtopics, this section focuses on the age difference in posting volumes per topic.
Results reported in this section (Figure \ref{fig:2013v2023ageYear}) are based on a segmented understanding of the posts' authors' ages, where those under 13 were categorized as "pre-teens", those aged 13-17 were "teens", and those 18 or older were "young adults". By comparing the distribution of the focal topics in 2013 to those in 2023 across age groups, we find that \textit{Family \& Parenting} rose significantly for pre-teens (21.1\% $\rightarrow$ 33.9\%). In direct contrast, \textit{Family \& Parenting} decreased slightly over the same period for teens (19\% $\rightarrow$ 17.8\%) and young adults (10.2\% $\rightarrow$ 10.1\%). The proportion of \textit{Mental Health \& Well-Being} to overall focal topic posts was more stable over time, with none of the age groups experiencing a change of magnitude 2.6\% or greater. Similar results were found for \textit{Relationships \& Friendships} for pre-teen and teens, with the former rising from 28.9\% to 29.4\% and the latter falling from 39.9\% to 39.3\%. For young adults, however, \textit{Relationships \& Friendships} declined sharply with a decrease from 43.8\% to 26.1\%. Regarding \textit{Body Image}, young adults were again the only age group to see a large change (15.9\% $\rightarrow$ 6\%), with pre-teens rising by 0.9\% (5.3\% $\rightarrow$ 6.4\%) and teens falling 0.6\% (9\% $\rightarrow$ 8.4\%). Finally, \textit{Sex \& Intimacy} showed the least consistent behavior from 2013 to 2023. While only a marginal increase was observed for teens (10.3\% $\rightarrow$ 11.7\%), a sharp decline was noted for pre-teens (21.1\% $\rightarrow$ 9.2\%) and an extreme increase was found for young adults (14.2\% $\rightarrow$ 42\%).

\begin{figure}[H]
    \centering
    \includegraphics[width = 0.8\linewidth]{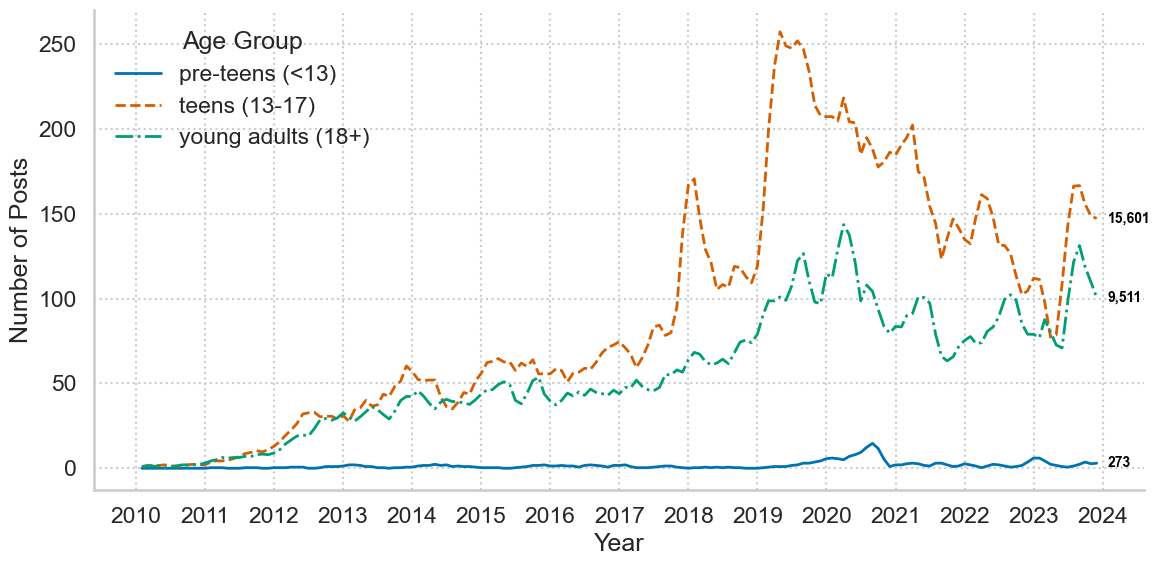}
    \caption{Relationship \& Friendship Topic Posting Volume over Time by Age Group (end values are cumulative totals)}
    \label{fig:relationAgeOverTime}
\end{figure}

Because the focal topics consist of multiple subtopics, we also analyzed the individual subtopics to better understand the reasons for the observed increases and decreases in their relative prominence. While \textit{relationships} consistently accounted for the lion's share of the \textit{Relationships \& Friendship} posts, the proportion declined (Figure \ref{fig:relationAgeOverTime}) across all age groups: pre-teens (90.9\% $\rightarrow$ 78.1\%), teens (80.6\% $\rightarrow$ 75.8\%), and young adults (77.8\% $\rightarrow$ 77.2\%). Within the \textit{Sex \& Intimacy} topic, the proportion of posts focusing on \textit{Porn \& Sexual Acts} declined in all age groups, with teens seeing a far smaller decrease of 5.9\% (61.9\% $\rightarrow$ 56\%) than pre-teens (62.5\% $\rightarrow$ 30\%) or young adults (58.3\% $\rightarrow$ 26.3\%). This was accompanied by comparable increases in posts addressing \textit{Sexual Kinks \& Roleplay}, where pre-teens rose 32.5\% (37.5\% $\rightarrow$ 70\%), teens increased 5.9\% (38.1\% $\rightarrow$ 44\%), and young adults climbed 42\% (41.7\% $\rightarrow$ 73.7\%). The subtopics of \textit{Mental Health \& Well-Being} behaved similarly across age groups, with \textit{Fear} consistently falling (pre-teens: 55.6\% $\rightarrow$ 34.8\%; teens: 47\% $\rightarrow$ 28\%; young adults: 48.3\% $\rightarrow$ 27.2\%) and \textit{Depression \& Poor Mental Health} rising (pre-teens: 44.4\% $\rightarrow$ 65.2\%; teens: 53\% $\rightarrow$ 72\%; young adults: 51.7\% $\rightarrow$ 72.8\%) over time. For the topics of \textit{Family \& Parenting} and \textit{Body Image}, the posts for both 2013 and 2023 consisted of the same singular subtopic, resulting in no significant changes across the decade.

Overall, the results show that changes in topic engagement were not uniform across age groups, with young adults exhibiting the largest shifts in posting distributions over time. Pre--teens and teens, by contrast, showed relatively stable patterns across most topics, with only modest changes in proportional representation. Notably, \textit{Mental Health \& Well-Being} remained comparatively stable across all groups, while \textit{Sex \& Intimacy} displayed the greatest divergence.

\subsection{Psycholinguistic Patterns and Variations among Topics}
\label{sec:liwc_analysis}


\begin{table}[h]
\scriptsize
\centering
\caption{Most Prominent Youth-Safety-Related Subtopics}
\label{tab:nine_prom_subtopics}
\begin{tabular}{l r}
\toprule
\textbf{Subtopic} & \textbf{Percentage of Posts (entire dataset)} \\
\midrule
\textit{Relationships} & 4.57\\
\textit{Parent-Child Abuse} & 2.42\\
\textit{Depression \& Poor Mental Health} & 2.24\\
\textit{Sexual Kinks \& Roleplay} & 1.92\\
\textit{Porn \& Sexual Acts} & 1.25\\
\textit{Weight Loss \& Body Shaming} & 1.21\\
\textit{Friendship} & 1.15\\
\textit{Fear} & 1.14\\
\bottomrule
\end{tabular}
\end{table}

Next, we investigate linguistic differences in youth authored posts across topics using methods described in Section \ref{sec:LIWC_method}. 
Per the official documentation \cite{liwcLIWCx2014, liwc_manual}, LIWC outputs primarily represent proportions of words belonging to specific linguistic categories (e.g., \texttt{emo\_pos}) relative to the total number of words in a text. Accordingly, while we understand the LIWC scores to indicate the psycholinguistic patterns exhibited by the various posts, we in no way interpret this as evidence of causal relationships between the topics and their linguistic patterns. That said, understanding what linguistic distinctions can be made between topics can aid in better identification of youth--authored discourse; as such, we examined which of focal topic pairs both statistically and practically differed on the basis of one or more LIWC categories (See Table~\ref{tab:focal_topics_cliff}). As noted in the Methods section (Section \ref{sec:LIWC_method}), we report only those pairs whose differences were significant and whose effect size magnitudes ($|\delta|$) were large (>0.47). 

Although 15 subtopic pairs differed significantly with regard to a given LIWC category, these pairs only in reference to 4 of the 7 other subtopics: \textit{Depression \& Poor Mental Health}, \textit{Friendship}, \textit{Parent--Child Abuse}, \textit{Relationships}. In what follows, we explain which LIWC categories were responsible for the separation of each subtopic pair and discuss briefly the significance of their respective effect sizes.

\textit{Depression \& Poor Mental Health} was the most diverse with regard to the subtopics from which it was linguistically distinct. In comparison to \textit{Sexual Kinks \& Roleplay}, \textit{Depression \& Poor Mental Health} exhibited greater usage of both negative emotion ($\delta = 0.6411$) and negative tone ($\delta = 0.5456$) language, suggesting that \textit{Depression \& Poor Mental Health} posts were typically framed in more affectively negative terms. \textit{Depression \& Poor Mental Health} also showed higher use of negative emotion language when compared to \textit{Weight Loss \& Body Shaming} ($\delta = 0.4763$), reflect a comparatively more ambivalent emotional tone in the latter subtopic. Notably, the depression subtopic also surpassed the \textit{Fear} subtopic in the use of \texttt{emotion} vocabulary ($\delta = 0.4753$). While other subtopics include integral emotional elements, the relatively lower use of emotion-related vocabulary in the latter subtopic suggests a more analytic or descriptive framing. Lastly, \textit{Parent--Child Abuse} saw greater use of third--person pronouns than \textit{Depression \& Poor Mental Health} ($\delta = -0.4861$), which may suggest that the latter is more self--focused.

\begin{table}[h]
\scriptsize
\centering
\caption{Focal topic pairs with significant differences and effect sizes (Cliff's Delta) of magnitude 0.47 or greater. All Holm--corrected p--values were less than 0.001. Positive values of $\delta$ indicate higher LIWC feature usage in Topic 1 relative to Topic 2, while negative values indicate higher usage in Topic 2. Cliff's $\delta$ ranges from $-1$ to $1$, with larger absolute values indicating stronger effects. Only comparisons meeting the threshold for large effects are shown.}
\label{tab:focal_topics_cliff}
\begin{tabular}{p{0.3\textwidth} p{0.3\textwidth} p{0.2\textwidth} p{0.1\textwidth}}
\toprule
\textbf{Topic 1} & \textbf{Topic 2} & \textbf{LIWC category} & \textbf{Cliff's $\delta$} \\
\midrule
Depression \& Poor Mental Health & Sexual Kinks \& Roleplay & emo\_neg & 0.641 \\
Parent--Child Abuse & Weight Loss \& Body Shaming & shehe & 0.572 \\
Depression \& Poor Mental Health & Sexual Kinks \& Roleplay & tone\_neg & 0.546 \\
Parent--Child Abuse & Porn \& Sexual Acts & shehe & 0.532 \\
Parent--Child Abuse & Weight Loss \& Body Shaming & ppron & 0.507 \\
Parent--Child Abuse & Sexual Kinks \& Roleplay & shehe & 0.504 \\
Relationships & Weight Loss \& Body Shaming & pronoun & 0.499 \\
Parent--Child Abuse & Sexual Kinks \& Roleplay & emo\_neg & 0.496 \\
Parent--Child Abuse & Weight Loss \& Body Shaming & pronoun & 0.494 \\
Relationships & Weight Loss \& Body Shaming & shehe & 0.489 \\
Depression \& Poor Mental Health & Weight Loss \& Body Shaming & emo\_neg & 0.476 \\
Depression \& Poor Mental Health & Fear & emotion & 0.475 \\
Depression \& Poor Mental Health & Parent--Child Abuse & shehe & -0.486 \\
Friendship & Relationships & shehe & -0.510 \\
Friendship & Parent--Child Abuse & shehe & -0.591 \\
\bottomrule
\end{tabular}
\end{table}

Following from this, \textit{Parent--Child Abuse} was further separated from 3 other subtopics: \textit{Weight Loss \& Body Shaming}, \textit{Porn \& Sexual Acts}, and \textit{Sexual Kinks \& Roleplay}. The first pair---\textit{Parent--Child Abuse} and \textit{Weight Loss \& Body Shaming}---differed with regard to 3 LIWC categories: \texttt{ppron} ($\delta = 0.5067$), which indicated the use of personal pronouns, \texttt{shehe} ($\delta = 0.4936$), which captures third--person pronouns, and \texttt{pronoun} ($\delta = 0.5723$), which represents overall use of pronouns. The greater use of pronouns could suggest a directive tone, with post authors addressing specific audiences rather than making general statements or asking broad questions. Similarly, the \texttt{shehe} LIWC category split \textit{Parent--Child Abuse} from \textit{Porn \& Sexual Acts} ($\delta = 0.5318$) and \textit{Sexual Kinks \& Roleplay} ($\delta = 0.5036$), again suggesting a stronger emphasis on others--directed post in the \textit{Parent--Child Abuse} subtopic. The abuse subtopic also contained more usage of \texttt{emo\_neg} LIWC vocabulary than \textit{Sexual Kinks \& Roleplay} ($\delta = 0.496$). Given the generally more supportive nature of the latter subtopic's posts, this is to be expected.

When we examined subtopic pairs including \textit{friendship}, we again found that \texttt{shehe} helped to distinguish subtopics. Both \textit{Relationships} ($\delta = -0.5101$) and \textit{Parent--Child Abuse} ($\delta = -0.5912$) surpassed \textit{Friendship} in their use of third--person singular pronouns. As friendships are fundamentally about intimate personal relationships, these differences in \texttt{shehe} vocabulary may represent the potential for more impersonal or abstract discourse found in posts from the \textit{Relationships} and \textit{Parent--Child Abuse} subtopics.

Finally, the \textit{Relationships} subtopic was differentiated from the \textit{Weight Loss \& Body Shaming} based on two LIWC categories: \texttt{pronoun} ($\delta = 0.4989$) and \texttt{shehe} ($\delta = 0.489$). While the \texttt{pronoun} category encompasses the \texttt{shehe} category, the presence of both as differentiating factors suggests that third--person singular pronouns are responsible for much of the influence coming from the larger category. The elevated use of pronouns in the \textit{Relationships} subtopic may be explained by the other subtopic including content that is directed toward general audiences (e.g., weight loss tip posts) instead of representing interpersonal interaction.

\section{Discussion}
\label{sec:discussion}
Section \ref{sec:discussion_rq1} examines how youth posting habits and discussion topics evolved from 2010 to 2023. Section \ref{sec:discussion_rq2} extends this by highlighting differences in posting volumes and content across pre-teens, teens, and young adults. Section \ref{sec:discussion_rq3} focuses on the linguistic properties of youth safety–related subtopics and underscores their interconnected nature.


\subsection{The Evolution of Youth-Discussed Topics on Reddit (RQ1)}
\label{sec:discussion_rq1}

The breadth and coherence of topics in our dataset offer confirmatory evidence for prior suggestions that Reddit facilitates disclosures that may be less palatable in identity--linked spaces \cite{kahlow2024beyond}. Our results show that youth not only use Reddit to discuss highly sensitive topics related to their physical and mental safety, but have done so consistently since before the modern ubiquity of social media took hold \cite{Pew2025TeensSocialMedia}. 
As digital natives and early adopters of social media \cite{Mertala2024DigitalNatives, Janschitz2022DigitalNatives}, youth are drawn to platforms like Reddit as venues for discussions spanning the innocuous to the socially taboo. By systematically identifying thousands of posts on child abuse, sexual predation, harassment, suicide, and related harms, we extend prior work on youth digital risks \cite{Livingstone2014Harms} by demonstrating their prevalence over time. Moreover, our longitudinal findings suggest that these risks have been enduring features of youths' digital experiences, even as their specific manifestations evolve. 
This positions Reddit as a useful proxy for youths' collective concerns and perceived dangers, complementing prior research that has largely relied on platform--specific or short--term analyses. 
Bearing in mind the strictness of our data cleaning process, which removed over 68\% of the initially collected data, the remaining dataset of over 440,000 posts suggests that Reddit may be even more topically diverse for youth than portrayed here. 
It is also significant that, despite Reddit's official rules requiring users to be a minimum of 13 years of age to use the platform and 18 to interact with NSFW subreddits \cite{redditRules, redditUserAgreement}, our data provides evidence that many underage persons engage in graphic or otherwise sensitive discussions.

Youth on Reddit are far from uniform in their behaviors and the topics with which they engage. As evidenced by the results of the Brown--Forsythe test (see Section \ref{sec:topics_analysis}), posting trends varied significantly across the 24 topics. However, while the frequency with which each topic rose or fell in prominence (“spikes”/“drops”) did not differ significantly between topics, the degree to which particular topics increased or decreased in their share of total youth--authored posts did vary meaningfully. This suggests that changes in topic--specific posting trends reflected broader shifts in overall youth posting volume rather than differences in the content distinguishing individual topics. This should not be taken to mean that topic--specific features played no role in shaping youth engagement over time. On the contrary, the finding that between 37\% and 43.8\% of the 276 possible topic pairs differed significantly in the magnitude or variance of their posting volume fluctuations indicates that youth devote substantially different amounts of attention to different topics. These findings align with those of \cite{mauri2021complementing, woodward2025time, pettyjohn2025m, Kumar2023Understanding, dan2025exploring}, as they have elsewhere shown that youth engage in both inter-- and intra--topic exploration. Moreover, the variability in how subtopics are distributed across subreddits within each topic is indicative of youth clustering around topics rather than particular subreddits, akin to the concept of Reddit communities proposed by \cite{Sawicki2023CrosspostNet}.

When restricted to our focal topics---\textit{"Crime, Law \& Sociopolitical Issues," "Relationships \& Friendships," "Sex \& Intimacy," "Mental Health \& Well-being," "Physical Health \& Medicine," "Body Image", "Family \& Parenting"}---these patterns become stark. In assessing the relative variances of our focal topics, we found that \textit{Body Image} and \textit{Family \& Parenting} had far smaller variances than the other topics, meaning that these two topics displayed more consistent trends. Given these two topics proximity to mental health and peer support \cite{Voelker2015BodyImage, TortNasarre2023PositiveBodyImage}, it is to be expected that youth would be more likely to engage in routine and protracted conversations when discussing either of these topics. Unlike the topic of \textit{Crime, Law \& Sociopolitical Issues}, which is predominantly event-driven \cite{Rasmussen2023EventDriven}, topics such as \textit{"Relationships \& Friendships," "Sex \& Intimacy," "Mental Health \& Well-being," "Physical Health \& Medicine,"} and \textit{"Body Image"} are more representative of an individual's psychosocial idiosyncracies \cite{Jaeger2014}. 

It is important to discuss the subtopics driving the extreme fluctuations in the posting volumes of each focal topic, as these signal larger, albeit sometimes temporary, shifts in the focus of a given topic. While direct causal relationships between real--world events and the shifts in subtopics cannot be established on the basis of our results alone, we highlight here some of the events that might have underpinned the large changes in the behavior of the focal topics. Although many of the topics are U.S.--centric, prior work has shown that approximately 35\% of Reddit engagement comes from U.S.--based users \cite{bozarth2023role}. Therefore, while we do not claim that our results or subsequent discussion capture the totality of youth-authored Reddit content, we regard them as representative of a substantial portion of it. Overall, we argue that many significant real--world events that affect youth offline are clearly reflected in their Reddit discourses.

\subsubsection{Crime, Law \& Sociopolitical Issues}
\label{sec:disc_crime}
The first extreme spike, occurring in 2012 and primarily driven by a rise in posts focusing on \textit{Religion} and \textit{Islam}, aligns with two major events that carried global effects: the assault on the U.S. consulate in Benghazi and the 2012 U.S. presidential election. Against the backdrop of an already rising focus on religion in U.S. political discourse and efforts to bolster youth engagement in the political process \cite{pew2012mediaReligionElection, pew2012religionNews2011, furman2012religionElection}, both events captured significant public attention and positioned religion as a topic of growing relevance to young adult voters. Thus, while not necessarily directly inducing the 2012 spike, it reflects a growing tide of religion being portrayed as a youth--involving subject. \textit{Religion} remained the largest subtopic throughout the 2018 and 2020 spikes, with a slightly smaller uptick in \textit{US politics} posts accompanying this shift. Events such as the upholding of then--President Trump’s travel bans against Muslim--majority nations by the U.S. Supreme Court \cite{MigrationPolicy2018TravelBan}, the 2018 Tree of Life Synagogue shooting in Pittsburgh, Pennsylvania \cite{Tobias2020TreeOfLife}, record--high youth turnouts in both the 2018 and 2020 U.S. elections \cite{harvard2018_memoYoungVoters, circle2019_28Youth2018, circle2021_halfYouth2020}, and the rise of progressive youth activism around immigration, racial justice, environmentalism, and LGBTQ+ rights \cite{Princeton2020YouthActivism} coincide with many of the more extreme fluctuations, suggesting a potential link between these events and youth discourse topic balance.

\subsubsection{Physical Health \& Medicine}
\label{sec:disc_physical}
Given the perpetual importance of a person's physical health, it is somewhat surprising that the posting volume of \textit{Physical Health \& Medicine} fluctuated as much as it did. With the exception of the extreme spike in 2020, when \textit{Vaccines \& Covid} surged to become the largest subtopic, it is difficult to ascertain the underlying events that may have motivated these extreme rises and falls in the topic's prominence. That said, we suggest that the decline of physical health--related posts may be more a result of the rise of mental health than an outright reduction in the perceived importance of physical health. In particular, in the early to mid-2010s, mental health gained prominence in popular culture, often being characterized as the central problem underpinning many physical ailments \cite{Stupinski2022}. This was echoed by celebrities who explicitly highlighted their lived experiences of physical and mental health challenges instigating or exacerbating each other \cite{allureLadyGaga}, a phenomenon that has elsewhere been shown to carry nontrivial impacts on shaping discourses around mental health \cite{Gronholm2022}.   

\subsubsection{Mental Health \& Well-Being}
\label{sec:disc_mental}
The discussion of \textit{Mental Health \& Well-Being} on Reddit is arguably more reflective of the larger shifts in how Reddit has been perceived and used over time than other topics. Early posts most frequently focused on \textit{Fear} and \textit{Dreams}, often referencing public hysteria events such as the abundance of end--of--world predictions in 2012 \cite{Ipsos2012EndWorld}. This aligns with much of Reddit's early content being satirical, comedic, or otherwise light-hearted \cite{Proferes2021RedditOverview}. By the time mental health became a cultural touchstone in the mid--2010s \cite{Stupinski2022}, Reddit had also evolved into a digital space that attracted large numbers of youth--led discussions (see Figure \ref{fig:postPerYEar}). Thus, the prevalence of conversations centered around \textit{Depression \& Poor Mental Health} that transpired on Reddit since 2016 serves as further evidence that youth-authored Reddit posts can, and often do, reflect broader cultural phenomena.

\subsubsection{Sex, Intimacy, and Body Image}
\label{sec:disc_sex}
Although the \textit{Sex \& Intimacy} and \textit{Body Image} topics display distinct trajectories, their respective trends appear driven by partially overlapping underlying dynamics. While not directly aligned, notions of what constitutes an attractive body type often resemble conversations about what is acceptable to discuss regarding personal sexual preferences. In particular, both sexuality and body image have, in recent years, become central topics in progressive social movements \cite{Russell2019SexualMinority} typically aimed at removing older and purportedly prejudicial conceptions of "normal" sex and bodies from public discourse. This has manifested in both highly visible environments, such as television series that increasingly glorify body positivity and feature overt depictions of sex \cite{Maes2023-ni}, as well as in the proliferation of fetish--specific dating apps \cite{airFeeldJustin}. In tandem with Reddit's established encouragement of candid conversation through the pseudonymous status of its users, it is unsurprising that many youth have increasingly utilized Reddit as an opportunity to disclose and discuss their perspectives on sex, intimacy, and body image. 

It should also noted that we removed posts that served only to redirect Reddit users to other platforms during the cleaning process. Many of these posts targeted digital sex sites such as OnlyFans, which has been argued to contribute to the normalization of \textit{sexual kinks \& roleplay} during the Covid-19 pandemic (2020--2022) \cite{Lippmann2023, Litam2022}. These potential links between youth--authored Reddit posts and adult--centered content suggest that the topics of \textit{Sex \& Intimacy} and \textit{Body Image} on Reddit may not be as tied to the platform's particular characteristics as other topics.

\subsubsection{Relationships \& Friendships}
\label{sec:disc_relationships}
As a topic composed of only two subtopics---\textit{Romantic/Sexual Relationships} and \textit{Friendships}---this topic has shown consistent and substantial growth in relative prominence across the full timeline of our dataset. While represented by a nontrivial number of posts, those focused on \textit{Friendship} totaled to only a quarter the number of posts discussing \textit{Relationships}. This is especially apparent beginning in 2017, when the \textit{Relationships \& Friendships} topic began to feature a large number of posts addressing users' opinions of online dating. Given the now abundant evidence that reviews on platforms like the Apple App Store and Google Play store often engage in censorship and manipulative designs \cite{Martens2019, Olson2024}, Reddit is often used as an alternative review platform \cite{YinSadowski2024}, especially when the apps being considered carry socially precarious associations.

Among the many significant shifts in the topic trends we discuss here, the most impactful may be the 2019 extreme spike in the volume of posts related to the \textit{Relationships} subtopic. Although posts authored at the time predate the popularization of phrases like "epidemic of loneliness" \cite{EClinicalMedicine2023} and "dating recession" \cite{Rosenfeld2025}, many of the warning signs of the now widely recognized decline in human--to--human romantic and sexual relationships were already evident in youths' posts by 2019. This decline did not begin in the 2010s; earlier research documented reductions of up to 33\% in the number of persons the average U.S. adult reported being able to converse about sensitive topics with \cite{McPherson2006}, as well as a sharp decline in sexual activity among U.S. adults aged 18--24 from 2000 to 2018 \cite{Ueda2020}. Our results therefore suggest that its impact on younger persons accelerated in the mid to late 2010s and persisted through the end of our dataset.

\subsubsection{Family \& Parenting}
\label{sec:disc_family}
Given that the \textit{Family \& Parenting} topic is predominantly about \textit{Parent--Child Abuse}, we suspect that changes in the topic’s prominence are largely responsive to highly publicized real--world scandals and domestic violence trends. Notable events in the years when the topic’s posting volume rose sharply (2012, 2023) include the conviction of Jerry Sandusky, a former coach at Penn State University, on 45 counts of child sexual abuse \cite{DailyCollegian2021}, revelations of the extent of child assaults in Catholic institutions \cite{USCCB2018}, and a steep uptick in domestic violence during the Covid--19 pandemic \cite{Piquero2021, UCDAvis2023}. Although these first two examples do not represent genuine parent--child dynamics, the involvement of father figures (e.g., coaches, priests) in the abuse of children may have inspired youth to disclose their own experiences of abuse on Reddit. More generally, this shows youth displaying a willingness to disclose their own accounts of abuse in public, albeit pseudonymous, settings.

\subsection{Reddit Topics across Different Youth Age Groups (RQ2)}
\label{sec:discussion_rq2}
While the insights discussed in Section \ref{sec:discussion_rq1} are themselves robust, they do not reveal the full story of how the topics discussed by youth on Reddit fluctuate across age groups and time. The heterogeneity of various age groups---pre--teens (<13), teens (13--17), young adults (18+)---with regard to youth safety--related topics is evidently nontrivial and therefore merits explicit attention. Note that we limited our analysis to a comparison between 2013 and 2023 in order to both ensure sufficient volumes of posts in each year and avoid biases resulting from overlapping world events.

It is important to note that the subtopic \textit{Parent--Child Abuse} includes both accounts of genuine abuse and complaints levied by youth of perceived unfair treatment. This is representative of youths' difficulties differentiating neglect from outright abuse \cite{Pediatrics2012}, a reality that is especially pronounced when youth are in their pre--teen years \cite{Lavoie2022}. Accordingly, our results showing a higher posting volume for the \textit{Family \& Parenting} topic among pre--teens than either of the other two age groups aligns well with established research. Moreover, this trend is evident in both 2013 and 2023, with the contrast between age groups growing over time. Whether this growing separation resulted from improved understanding of what does and does not constitute abuse among older youths or from a genuine increase in the volume of abuse directed toward pre-teens cannot be determined from our data alone, but the aforementioned evidence of increased frequency of domestic abuse \cite{Piquero2021, UCDAvis2023} (see Section \ref{sec:disc_family}) lends credence to the latter interpretation.

Contrary to what might be assumed given the growing public focus on mental health \cite{DomingoEspineira2024}, \textit{Mental Health \& Well-Being} saw only minimal shifts between 2013 and 2023, regardless of age group. Accounting for this discrepancy may be explained by considering the growing representation of mental health topics on other social media platforms. As evidenced by \cite{Marciano2022, Naslund2020, Issaka2024, Pavlova2020}, platforms like Twitter (X) and Instagram have become prominent venues for discussing one's mental health. Consequently, our results suggest not that the reported rise in attention dedicated to mental health is inauthentic, but that youth may have diversified which social media platforms they use to converse about mental health.

A similar cultural--data alignment applies to the \textit{Relationships} subtopic. Though the subtopic (and its parent topic \textit{Relationships \& Friendships}) remained stable from 2013 to 2023 for pre--teen and teenage youth, a drastic reduction in posting volume occurred for young adults over the decade. Consistent with evidence that the rise of short--term relationships is not confined to any particular region, having been documented in the U.K. \cite{Lim2024}, Israel \cite{Shulman2023}, and Uganda \cite{Choudhry2022}, among other regions, young adults' interpretations of relationships increasingly center on transactional, often sexual, dynamics. This interpretation is further supported by our finding that \textit{Sex \& Intimacy} increased in prominence among young adults from 14.2\% to 42\% of posts over the same period, while teens and pre-teens exhibited only minor fluctuations.

\subsection{Interpretations of the Linguistic Qualities of Prominent Youth Safety-Related Subtopics (RQ3)}
\label{sec:discussion_rq3}

The psycholinguistic distinctions observed across youth safety--related subtopics align with prior work demonstrating that linguistic features on Reddit reflect underlying differences in emotional expression and social orientation. However, rather than being broadly distributed across all subtopics, these distinctions are concentrated within a small subset---most notably \textit{Depression \& Poor Mental Health}, \textit{Parent--Child Abuse}, and relational domains such as \textit{Relationships} and \textit{Friendship}. This concentration suggests that linguistic variation in youth discourse is structured around a limited number of recurring dimensions.

One such dimension is affective expression. Prior LIWC--based studies have shown that mental health discourse is characterized by elevated emotional and affective language \cite{Kim2023MentalHealthReddit, ireland-etal-2023-sadness}, a pattern represented by the consistent separation of \textit{Depression \& Poor Mental Health} from other subtopics. This separation is driven not simply by the presence of emotional language, but by the relative intensity and persistence of negative affect, as evidenced by large--effect differences in both \texttt{emo\_neg} and \texttt{tone\_neg}. Notably, the distinction between \textit{Depression \& Poor Mental Health} and \textit{Fear} suggests that even within emotionally salient domains, linguistic expression varies in form. Whereas fear--related posts may reflect situational concerns, depression--related posts appear more consistently organized around sustained affective framing. This is consistent with prior findings that Reddit facilitates self--disclosure and emotionally expressive communication in mental health contexts \cite{de2014mental}, particularly under conditions of pseudonymity.

A second dimension, which appears more pervasively across subtopics, concerns interpersonal reference. The repeated role of pronoun--based LIWC categories---especially \texttt{shehe}---in distinguishing \textit{Parent--Child Abuse}, \textit{Relationships}, and \textit{Friendship} is consistent with prior research emphasizing the social and relational nature of youth discourse online \cite{pettyjohn2025m, sit2024youth}. Elevated use of third--person pronouns in these subtopics suggests that posts are frequently structured around interactions involving specific others, such as parents, partners, or peers. This aligns with qualitative findings that youth often use Reddit to narrate interpersonal experiences, seek advice, and interpret social situations, particularly in domains related to relationships and abuse \cite{carym2026, pettyjohn2025m}. In contrast, subtopics such as \textit{Weight Loss \& Body Shaming}, which exhibit comparatively lower pronoun usage, appear more consistent with generalized or advice--oriented discourse \cite{saily2011variation}, reflecting a different mode of engagement.

Importantly, these dimensions do not operate independently. Subtopics such as \textit{Parent--Child Abuse} are distinguished by both heightened interpersonal reference and, in some comparisons, increased negative emotional language, indicating that certain domains of discourse are both interpersonally grounded and affectively salient. This sort of interwoven linguistic structure generally conforms with prior literature on youth help--seeking behaviors, which highlight the interplay between emotional distress and interpersonal context in online disclosures \cite{sit2024youth}. That said, the absence of large--effect differences across many other subtopic pairs suggests that not all youth safety--related discussions are linguistically distinct at this level of analysis. Instead, psycholinguistic differentiation appears to be driven by those domains in which either emotional intensity or interpersonal relations are integral to how experiences are articulated. Our findings therefore reinforce prior work demonstrating the utility of LIWC for identifying meaningful linguistic variation in social media discourse \cite{melanson2025linguistic}, while also establishing that such variation is not uniform across youth--discussed topics.

\section{Limitations and Future Work}

Despite the volume of data considered for this study, several nontrivial limitations remain. First, our dataset slightly overestimates the ratio of male to female Reddit users: while the late--2024 balance was 59.1 to 39.1 (with 1.8\% unknown) \cite{statistaGlobalReddit}, our dataset shows 62.6 to 37.4, an overrepresentation of males by about 3.5\%. Second, we did not analyze the comments on youths’ posts. Given the dataset’s scale and topical diversity, we believe the insights reported here are of real consequence. Still, future work should examine the reactions and extended discourses captured in comment threads.

A portion of our discussion (see Sections \ref{sec:disc_crime}--\ref{sec:disc_family}) centered on the alignment of Reddit topics with real--world events. Given Reddit’s relatively recent rise to international prominence, we chose to limit our examples to events with clear impacts in countries where Reddit was already popular or to those with global effects. While we were cognizant of the U.S.-- and Eurocentric nature of many of these examples and made efforts to consider non--Western perspectives, a broader scope of world events would strengthen the analysis. 

Lastly, BERTopic, while a now common and well--regarded practice in HCI research, is far from the only reasonable method for performing large--scale topic modeling. Other researchers have highlighted the vulnerability of BERTopic to platform--specific biases \cite{koo-etal-2024-platform}, developing and promoting alternative models like \cite{platform_invariant_topic_modeling_repo}. Thus, if our findings are to be expanded beyond Reddit, the differences between various topic modeling algorithms should be explicitly examined. 

%

\section{Ethical Considerations}
\label{sec:ethical_considerations}
Given that our research focuses on vulnerable populations and discusses highly sensitive topics at times, we follow strict ethics--oriented protocols when designing the research methods and reporting our results. In earlier stages of data preparation, we leaned towards more conservative measures while inferring age and gender information from Reddit posts. Although prior scholarship demonstrates linguistic approaches to demographic inference on social media, this study relies exclusively on explicit self--disclosures to minimize bias and avoid misrepresentation. The data used in all research questions is public, and all posts deleted by the authors have been removed. To reduce potential risks, results are reported solely in high--level aggregate form, without drawing attention to any specific groups. Direct quotations are lightly rephrased to further reduce traceability, and the dataset itself will not be released publicly, adhering to Reddit's policies.
\section{Conclusion}

Our study demonstrates that youth--authored Reddit posts provide a valuable lens for understanding long--term shifts in online safety--related discourse. By combining topic modeling with linguistic analysis, we showed that topics traveled distinct trajectories, varied across age groups, and reflected broader sociocultural dynamics. These results highlight the importance of considering both age and context when interpreting online conversations, particularly for issues that intersect with youth well--being. Collectively, these insights underscore the potential of youth--authored data to reshape how researchers, designers, and policymakers approach online safety.

\vspace{0.5cm}
\noindent \textbf{AI Usage Disclosure:} GPT-5.2 was used to resolve minor grammatical errors and improve table formatting. Google Scholar Labs was used to identify some of the supporting literature. 


 



\section{Acknowledgments}
Left blank for peer review.

\setlength{\bibsep}{0.5\baselineskip}
\bibliography{references}

@misc{Saini2025RedditStatistics,
  author       = {Manisha Saini},
  title        = {50+ Key Reddit Statistics You Must Know in 2025},
  howpublished = {Blog post on Cropink},
  year         = {2025},
  month        = {mar},
  note         = {Reviewed by Leszek Dudkiewicz},
  url          = {https://cropink.com/reddit-statistics?}
}

@misc{BusinessOfApps2025RedditStatistics,
  author       = {{Business of Apps}},
  title        = {Reddit Statistics},
  howpublished = {Data page on Business of Apps},
  year         = {2025},
  note         = {Accessed via URL with UTM source},
  url          = {https://www.businessofapps.com/data/reddit-statistics/?}
}

@article{zhang2021teens,
  title        = {Teens' Social Media Engagement during the COVID-19 Pandemic: A Time Series Examination of Posting and Emotion on Reddit},
  author       = {Zhang, Shuya and Liu, Mengqi and Li, Yuchen and Chung, Ji Eun},
  journal      = {International Journal of Environmental Research and Public Health},
  volume       = {18},
  number       = {19},
  pages        = {10079},
  year         = {2021},
  publisher    = {MDPI},
  doi          = {10.3390/ijerph181910079},
  url          = {https://doi.org/10.3390/ijerph181910079}
}

@article{woodward2025time,
  author       = {Woodward, M. J. and McGettrick, C. R. and Dick, O. G. and others},
  title        = {Time Spent on Social Media and Associations with Mental Health in Young Adults: Examining TikTok, Twitter, Instagram, Facebook, Youtube, Snapchat, and Reddit},
  journal      = {Journal of Technology in Behavioral Science},
  year         = {2025},
  publisher    = {Springer},
  doi          = {10.1007/s41347-024-00474-y},
  url          = {https://doi.org/10.1007/s41347-024-00474-y}
}

@inproceedings{mauri2021complementing,
author = {Mauri, Andrea and Psyllidis, Achilleas and Bozzon, Alessandro and Lee, Ju-Sung and Pridmore, Jason and van Zoonen, Liesbet and Giest, Sarah},
title = {Complementing Studies on Vulnerable Youths with Reddit Data},
year = {2021},
isbn = {9781450389778},
publisher = {Association for Computing Machinery},
address = {New York, NY, USA},
url = {https://doi.org/10.1145/3464385.3464703},
doi = {10.1145/3464385.3464703},
booktitle = {Proceedings of the 14th Biannual Conference of the Italian SIGCHI Chapter},
articleno = {5},
numpages = {8},
location = {Bolzano, Italy},
series = {CHItaly '21}
}

@article{kahlow2024beyond,
  author  = {Kahlow, Emily J.},
  title   = {Beyond the Surface: Reddit’s Anonymity Facilitates Deeper Disclosure Compared to Facebook},
  journal = {International Journal of Cyber Behavior, Psychology and Learning},
  volume  = {14},
  number  = {1},
  year    = {2024},
  doi     = {10.4018/IJCBPL.343629},
  url     = {https://doi.org/10.4018/IJCBPL.343629}
}

@inproceedings{rybnicek2013facebook,
  author    = {Marlies Rybnicek and Rainer Poisel and Simon Tjoa},
  title     = {Facebook Watchdog: A Research Agenda for Detecting Online Grooming and Bullying Activities},
  booktitle = {2013 IEEE International Conference on Systems, Man, and Cybernetics (SMC)},
  year      = {2013},
  pages     = {2854--2859},
  publisher = {IEEE},
  doi       = {10.1109/SMC.2013.487}
}

@article{tang2024social,
  author  = {Ruizi Tang and Ruiyang Shi and Yixiao Zhao and Yuer Chen and Boao Hua},
  title   = {Social Media Use Effects on Self-Cognition and Behavior Patterns under the Network Environment},
  journal = {Open Journal of Social Sciences},
  year    = {2024},
  volume  = {12},
  number  = {3},
  pages   = {360--381},
  doi     = {10.4236/jss.2024.123025}
}

@misc{reddit_transparency_2023b,
  title        = {Transparency Report: July to December 2023},
  author       = {{Reddit Inc.}},
  howpublished = {\url{https://redditinc.com/policies/transparency-report-july-to-december-2023}},
  year         = {2024},
  note         = {Accessed: 2025-09-05}
}

@inproceedings{simhadri2023csa,
  author    = {Siva Sahitya Simhadri and Tatiana R. Ringenberg},
  title     = {Child Sexual Abuse Awareness and Support Seeking on {Reddit}: A Thematic Analysis},
  booktitle = {Companion Proceedings of the ACM Web Conference 2023},
  year      = {2023},
  publisher = {ACM},
  doi       = {10.1145/3543873.3587363}
}

@article{janaswamy2024exploring,
  title={Exploring Climate Change Discourse: Measurements and Analysis of Reddit Data},
  author={Janaswamy, Smriti and Blackburn, Jeremy},
  journal={arXiv preprint arXiv:2412.01111},
  year={2024}
}

@article{johnson2022sexually,
  title={Sexually Transmitted Disease--Related Reddit Posts During the COVID-19 Pandemic: Latent Dirichlet Allocation Analysis},
  author={Johnson, Amy K and Bhaumik, Runa and Nandi, Debarghya and Roy, Abhishikta and Mehta, Supriya D},
  journal={Journal of medical Internet research},
  volume={24},
  number={10},
  pages={e37258},
  year={2022},
  publisher={JMIR Publications Toronto, Canada}
}

@article{valensise2021entropy,
  title={Entropy and complexity unveil the landscape of memes evolution},
  author={Valensise, Carlo M and Serra, Alessandra and Galeazzi, Alessandro and Etta, Gabriele and Cinelli, Matteo and Quattrociocchi, Walter},
  journal={Scientific Reports},
  volume={11},
  number={1},
  pages={20022},
  year={2021},
  publisher={Nature Publishing Group UK London}
}

@article{ford2023competition,
  title={Competition dynamics in the meme ecosystem},
  author={Ford, Trenton W and Krohn, Rachel and Weninger, Tim},
  journal={ACM Transactions on Social Computing},
  volume={6},
  number={3-4},
  pages={1--19},
  year={2023},
  publisher={ACM New York, NY}
}

@article{razi2023Sliding,
author = {Razi, Afsaneh and Alsoubai, Ashwaq and Kim, Seunghyun and Ali, Shiza and Stringhini, Gianluca and De Choudhury, Munmun and Wisniewski, Pamela J.},
title = {Sliding into My DMs: Detecting Uncomfortable or Unsafe Sexual Risk Experiences within Instagram Direct Messages Grounded in the Perspective of Youth},
year = {2023},
issue_date = {April 2023},
publisher = {Association for Computing Machinery},
address = {New York, NY, USA},
volume = {7},
number = {CSCW1},
url = {https://doi.org/10.1145/3579522},
doi = {10.1145/3579522},
journal = {Proc. ACM Hum.-Comput. Interact.},
month = apr,
articleno = {89},
numpages = {29}
}

@inproceedings{Razi2022Understanding,
author = {Ali, Shiza and Razi, Afsaneh and Kim, Seunghyun and Alsoubai, Ashwaq and Gracie, Joshua and De Choudhury, Munmun and Wisniewski, Pamela J. and Stringhini, Gianluca},
title = {Understanding the Digital Lives of Youth: Analyzing Media Shared within Safe Versus Unsafe Private Conversations on Instagram},
year = {2022},
isbn = {9781450391573},
publisher = {Association for Computing Machinery},
address = {New York, NY, USA},
url = {https://doi.org/10.1145/3491102.3501969},
doi = {10.1145/3491102.3501969},
booktitle = {Proceedings of the 2022 CHI Conference on Human Factors in Computing Systems},
articleno = {148},
numpages = {14},
location = {New Orleans, LA, USA},
series = {CHI '22}
}

@article{Alluhidan2024Teen,
author = {Alluhidan, Abdulmalik and Akter, Mamtaj and Alsoubai, Ashwaq and Park, Jinkyung Katie and Wisniewski, Pamela},
title = {Teen Talk: The Good, the Bad, and the Neutral of Adolescent Social Media Use},
year = {2024},
issue_date = {November 2024},
publisher = {Association for Computing Machinery},
address = {New York, NY, USA},
volume = {8},
number = {CSCW2},
url = {https://doi.org/10.1145/3686961},
doi = {10.1145/3686961},
journal = {Proc. ACM Hum.-Comput. Interact.},
month = nov,
articleno = {422},
numpages = {35}
}

@misc{Pew2025TeensSocialMedia,
  author       = {{Pew Research Center}},
  title        = {Teens and Social Media Fact Sheet},
  howpublished = {\url{https://www.pewresearch.org/internet/fact-sheet/teens-and-social-media-fact-sheet/}},
  month        = jul,
  year         = {2025},
  note         = {Accessed: 2025-09-09},
}

@inproceedings{Fischer2025Building,
author = {Fischer, Katrin and Xie, Louise and Arnold, Lauren Jade and Stevens, Robin},
title = {Building a Better Social Media Platform: Can We Codesign With Equity in Mind?},
year = {2025},
isbn = {9798400713941},
publisher = {Association for Computing Machinery},
address = {New York, NY, USA},
url = {https://doi.org/10.1145/3706598.3713901},
doi = {10.1145/3706598.3713901},
booktitle = {Proceedings of the 2025 CHI Conference on Human Factors in Computing Systems},
articleno = {953},
numpages = {10},
location = {
},
series = {CHI '25}
}

@ARTICLE{Sawicki2023CrosspostNet,
  title     = "Reddit {CrosspostNet---studying} Reddit communities with
               large-scale crosspost graph networks",
  author    = "Sawicki, Jan and Ganzha, Maria and Paprzycki, Marcin and
               Watanobe, Yutaka",
  journal   = "Algorithms",
  publisher = "MDPI AG",
  volume    =  16,
  number    =  9,
  pages     = "424",
  month     =  sep,
  year      =  2023,
  copyright = "https://creativecommons.org/licenses/by/4.0/",
  language  = "en"
}

@ARTICLE{Medvedev2018modelling,
  title        = "Modelling structure and predicting dynamics of discussion
                  threads in online boards",
  author       = "Medvedev, Alexey N and Delvenne, Jean-Charles and Lambiotte,
                  Renaud",
  year         =  2018,
  primaryClass = "cs.SI",
  eprint       = "1801.10082"
}

@misc{statistaGlobalReddit,
	author = {},
	title = {Distribution of Reddit users worldwide as of the second quarter of 2025, by gender},
	howpublished = {\url{https://www.statista.com/statistics/1255182/distribution-of-users-on-reddit-worldwide-gender/}},
	year = {2024},
	note = {[Accessed 09-09-2025]},
}

@inproceedings{shelby2023sociotechnical,
author = {Shelby, Renee and Rismani, Shalaleh and Henne, Kathryn and Moon, AJung and Rostamzadeh, Negar and Nicholas, Paul and Yilla-Akbari, N'Mah and Gallegos, Jess and Smart, Andrew and Garcia, Emilio and Virk, Gurleen},
title = {Sociotechnical Harms of Algorithmic Systems: Scoping a Taxonomy for Harm Reduction},
year = {2023},
isbn = {9798400702310},
publisher = {Association for Computing Machinery},
address = {New York, NY, USA},
url = {https://doi.org/10.1145/3600211.3604673},
doi = {10.1145/3600211.3604673},
booktitle = {Proceedings of the 2023 AAAI/ACM Conference on AI, Ethics, and Society},
pages = {723–741},
numpages = {19},
location = {Montr\'{e}al, QC, Canada},
series = {AIES '23}
}

@article{Nova2021facebook,
author = {Nova, Fayika Farhat and DeVito, Michael Ann and Saha, Pratyasha and Rashid, Kazi Shohanur and Roy Turzo, Shashwata and Afrin, Sadia and Guha, Shion},
title = {"Facebook Promotes More Harassment": Social Media Ecosystem, Skill and Marginalized Hijra Identity in Bangladesh},
year = {2021},
issue_date = {April 2021},
publisher = {Association for Computing Machinery},
address = {New York, NY, USA},
volume = {5},
number = {CSCW1},
url = {https://doi.org/10.1145/3449231},
doi = {10.1145/3449231},
journal = {Proc. ACM Hum.-Comput. Interact.},
month = apr,
articleno = {157},
numpages = {35}
}

@INPROCEEDINGS{thomas2021sok,
  author={Thomas, Kurt and Akhawe, Devdatta and Bailey, Michael and Boneh, Dan and Bursztein, Elie and Consolvo, Sunny and Dell, Nicola and Durumeric, Zakir and Kelley, Patrick Gage and Kumar, Deepak and McCoy, Damon and Meiklejohn, Sarah and Ristenpart, Thomas and Stringhini, Gianluca},
  booktitle={2021 IEEE Symposium on Security and Privacy (SP)}, 
  title={SoK: Hate, Harassment, and the Changing Landscape of Online Abuse}, 
  year={2021},
  volume={},
  number={},
  pages={247-267},
  doi={10.1109/SP40001.2021.00028}}

@inproceedings{medina2024ALWE,
author = {Medina, Jessica Y. and Young, Jordyn and Miller, Wendy Trueblood and Razi, Afsaneh},
title = {Exploring Online Support Needs of Adolescents Living with Epilepsy},
year = {2024},
isbn = {9798400711145},
publisher = {Association for Computing Machinery},
address = {New York, NY, USA},
url = {https://doi.org/10.1145/3678884.3681906},
doi = {10.1145/3678884.3681906},
booktitle = {Companion Publication of the 2024 Conference on Computer-Supported Cooperative Work and Social Computing},
pages = {558–564},
numpages = {7},
location = {San Jose, Costa Rica},
series = {CSCW Companion '24}
}

@inproceedings{de2014mental,
  title={Mental health discourse on reddit: Self-disclosure, social support, and anonymity},
  author={De Choudhury, Munmun and De, Sushovan},
  booktitle={Proceedings of the international AAAI conference on web and social media},
  volume={8},
  number={1},
  pages={71--80},
  year={2014}
}

@article{dan2025exploring,
  title={Exploring Suicide Factors in Online Discourse: Sentiment and Thematic Analysis of Reddit},
  author={Dan, Evan and Zhu, Jianfeng and Jin, Ruoming},
  journal={ACM Transactions on the Web},
  year={2025},
  publisher={ACM New York, NY}
}

@article{melanson2025linguistic,
  title={Linguistic Markers of Faster Life History Strategies among Teen Parents on Reddit: A Comparative Analysis},
  author={Melanson-Ricciardone, Sophia},
  journal={Discover Psychology},
  year={2025}
}

@article{pettyjohn2025m,
  title={“I’m not experienced… please send advice”: teens seeking information and advice about sexual behaviors on Reddit},
  author={PettyJohn, Morgan E and Cary, Kyla M and Nolen, Erin and Gallegos, Toni A},
  journal={The Journal of Sex Research},
  pages={1--12},
  year={2025},
  publisher={Taylor \& Francis}
}

@article{sit2024youth,
  title={Youth mental health help-seeking information needs and experiences: a thematic analysis of Reddit posts},
  author={Sit, Meghan and Elliott, Sarah A and Wright, Kelsey S and Scott, Shannon D and Hartling, Lisa},
  journal={Youth \& Society},
  volume={56},
  number={1},
  pages={24--41},
  year={2024},
  publisher={SAGE Publications Sage CA: Los Angeles, CA}
}

@misc{liwcLIWCx2014,
	author = {},
	title = {{LIWC} --- {LIWC} Analysis},
	url = {https://www.liwc.app/help/liwc#Interpreting-Output},
	year = {2014},
	note = {[Accessed 10-09-2025]},
}

@misc{liwc_manual, 
    title={The development and psychometric properties of LIWC-22}, 
    url={https://www.liwc.app/static/documents/LIWC-22%20Manual%20-%20Development%20and%20Psychometrics.pdf}, 
    journal={LIWC}, 
    publisher={Pennebaker Conglomerates, Inc.},
    author={Boyd, Ryan L. and Ashokkumar, Ashwini and Seraj, Sarah and Pennebaker, James W.},
    year = {2022}
}

@misc{redditUserAgreement,
  author       = {{Reddit Inc.}},
  title        = {Reddit User Agreement},
  year         = {2025},
  howpublished = {\url{https://redditinc.com/policies/user-agreement\#p\_37}},
  note         = {Accessed: 2025-09-11}
}

@misc{redditRules,
  author       = {{Reddit Inc.}},
  title        = {Reddit Content Policy (Reddit Rules)},
  howpublished = {\url{https://redditinc.com/policies/reddit-rules}},
  note         = {Accessed: 2025-09-11}
}

@misc{pew2012mediaReligionElection,
  author       = {{Pew Research Center}},
  title        = {The Media, Religion and the 2012 Campaign for President},
  year         = {2012},
  howpublished = {\url{https://www.pewresearch.org/journalism/2012/12/14/media-religion-and-2012-campaign-president/}},
  note         = {[Accessed: 2025-09-11]}
}

@misc{pew2012religionNews2011,
  author       = {{Pew Research Center}},
  title        = {Religion in the News: Islam and Politics Dominate Religion Coverage in 2011},
  year         = {2012},
  howpublished = {\url{https://www.pewresearch.org/religion/2012/02/23/religion-in-the-news-islam-and-politics-dominate-religion-coverage-in-2011/}},
  note         = {[Accessed: 2025-09-11]}
}

@misc{furman2012religionElection,
  author       = {{Furman University}},
  title        = {Religion and Republicans: The Presidential Election of 2012},
  year         = {2012},
  howpublished = {\url{https://www.furman.edu/wp-content/uploads/2019/03/Religion-in-the-2012-Presidential-Campaign.pdf}},
  note         = {[Accessed: 2025-09-11]}
}

@misc{circle2021_halfYouth2020,
  author       = {{Center for Information \& Research on Civic Learning and Engagement (CIRCLE), Tufts University}},
  title        = {Half of Youth Voted in 2020, An 11-Point Increase from 2016},
  year         = {2021},
  howpublished = {\url{https://circle.tufts.edu/latest-research/half-youth-voted-2020-11-point-increase-2016}},
  note         = {Accessed: 2025-09-11}
}

@misc{circle2019_28Youth2018,
  author       = {{Center for Information \& Research on Civic Learning and Engagement (CIRCLE), Tufts University}},
  title        = {28\% of Young People Voted in 2018},
  year         = {2019},
  howpublished = {\url{https://circle.tufts.edu/latest-research/28-young-people-voted-2018}},
  note         = {Accessed: 2025-09-11}
}

@misc{harvard2018_memoYoungVoters,
  author       = {{Institute of Politics, Harvard Kennedy School}},
  title        = {MEMO: Historic Turnout and Performance by Young Voters},
  year         = {2018},
  howpublished = {\url{https://iop.harvard.edu/press-releases/memo-historic-turnout-and-performance-young-voters}},
  note         = {Accessed: 2025-09-11}
}

@inproceedings{koo-etal-2024-platform,
  title     = {Platform-Invariant Topic Modeling via Contrastive Learning to Mitigate Platform-Induced Bias},
  author    = {Koo, Minseo and Kim, Doeun and Han, Sungwon and Park, Sungkyu Shaun},
  booktitle = {Findings of the Association for Computational Linguistics: EMNLP 2024},
  month     = nov,
  year      = {2024},
  address   = {Miami, Florida, USA},
  publisher = {Association for Computational Linguistics},
  pages     = {11123--11139},
  url       = {https://aclanthology.org/2024.findings-emnlp.650/},
  doi       = {10.18653/v1/2024.findings-emnlp.650}
}

@misc{platform_invariant_topic_modeling_repo,
  author       = {kde9867},
  title        = {Platform-Invariant Topic Modeling via Contrastive Learning},
  howpublished = {\url{https://github.com/kde9867/Platform-Invariant-Topic-Modeling}},
  year         = {2024},
  note         = {Accessed: 2025-09-11}
}

@article{Stupinski2022,
  author    = {Stupinski, Alexander M. and Alshaabi, Taha and Arnold, Michael V. and Adams, Jackson L. and Minot, Joshua R. and Price, Mackenzie and Dodds, Peter Sheridan and Danforth, Christopher M.},
  title     = {Quantifying Changes in the Language Used Around Mental Health on Twitter Over 10 Years: Observational Study},
  journal   = {JMIR Mental Health},
  year      = {2022},
  volume    = {9},
  number    = {3},
  pages     = {e33685},
  doi       = {10.2196/33685}
}

@article{Gronholm2022,
  author    = {Gronholm, Petra C. and Thornicroft, Graham},
  title     = {Impact of celebrity disclosure on mental health-related stigma},
  journal   = {Epidemiology and Psychiatric Sciences},
  year      = {2022},
  volume    = {31},
  pages     = {e62},
  doi       = {10.1017/S2045796022000488}
}

@misc{allureLadyGaga,
	author = {Andrea Park},
	title = {{L}ady {G}aga {O}pened {U}p {A}bout {S}truggling {W}ith {C}hronic {P}ain in {T}his {M}onth's {V}ogue --- allure.com},
	howpublished = {\url{https://www.allure.com/story/lady-gaga-vogue-fibromyalgia-chronic-pain-ptsd?}},
	year = {2018},
	note = {[Accessed 11-09-2025]},
}

@ARTICLE{Maes2023-ni,
  title     = "``I love my body; {I} love it all'' : Body positivity messages
               in youth-oriented television series",
  author    = "Maes, Chelly and Vandenbosch, Laura",
  journal   = "Mass Commun. Soc.",
  publisher = "Informa UK Limited",
  volume    =  26,
  number    =  1,
  pages     = "122--146",
  month     =  jan,
  year      =  2023,
  language  = "en"
}

@misc{airFeeldJustin,
	author = {Justin Lehmiller},
	title = {{F}eeld x {J}ustin {L}ehmiller, {K}insey {I}nstitute 2024 {R}eport / {F}eeld / {A}ir --- app.air.inc},
	howpublished = {\url{https://app.air.inc/a/cx2MUPVP0}},
	year = {2024},
	note = {[Accessed 11-09-2025]},
}

@article{Lippmann2023,
  author    = {Lippmann, Miriam and Lawlor, Nicole and Leistner, Christian E.},
  title     = {Learning on OnlyFans: User Perspectives on Knowledge and Skills Acquired on the Platform},
  journal   = {Sexuality \& Culture},
  year      = {2023},
  pages     = {1--21},
  doi       = {10.1007/s12119-022-10060-0},
  note      = {Advance online publication}
}

@article{Litam2022,
  author    = {Litam, Seo D. A. and Speciale, Melissa and Balkin, Richard S.},
  title     = {Sexual Attitudes and Characteristics of OnlyFans Users},
  journal   = {Archives of Sexual Behavior},
  year      = {2022},
  volume    = {51},
  number    = {6},
  pages     = {3093--3103},
  doi       = {10.1007/s10508-022-02329-0}
}

@article{Olson2024,
  author    = {Lauren Olson and Neelam Tjikhoeri and Emitzá Guzmán},
  title     = {The Best Ends by the Best Means: Ethical Concerns in App Reviews},
  journal   = {arXiv preprint arXiv:2401.11063},
  year      = {2024},
  doi       = {10.48550/arXiv.2401.11063}
}

@article{Martens2019,
  author    = {Daniel Martens and Walid Maalej},
  title     = {Towards Understanding and Detecting Fake Reviews in App Stores},
  journal   = {arXiv preprint arXiv:1904.12607},
  year      = {2019},
  doi       = {10.48550/arXiv.1904.12607}
}

@article{YinSadowski2024,
  author    = {Qian Yin and Sebastian Sadowski},
  title     = {A tale of two platforms: A comparative analysis of language use in consumer complaints on Reddit and Spotify Community},
  journal   = {Journal of Consumer Behaviour},
  year      = {2024},
  volume    = {23},
  number    = {4},
  pages     = {2071--2086},
  doi       = {10.1002/cb.2325}
}

@article{McPherson2006,
  author    = {McPherson, Miller and Smith-Lovin, Lynn and Brashears, Matthew E.},
  title     = {Social Isolation in America: Changes in Core Discussion Networks over Two Decades},
  journal   = {American Sociological Review},
  year      = {2006},
  volume    = {71},
  number    = {3},
  pages     = {353--375},
  doi       = {10.1177/000312240607100301},
  url       = {https://doi.org/10.1177/000312240607100301}
}

@article{Ueda2020,
  author    = {Ueda, Peter and Mercer, Catherine H. and Ghaznavi, Cyrus and Herbenick, Debby},
  title     = {Trends in Frequency of Sexual Activity and Number of Sexual Partners Among Adults Aged 18 to 44 Years in the US, 2000–2018},
  journal   = {JAMA Network Open},
  year      = {2020},
  volume    = {3},
  number    = {6},
  pages     = {e203833},
  doi       = {10.1001/jamanetworkopen.2020.3833},
  url       = {https://doi.org/10.1001/jamanetworkopen.2020.3833}
}

@article{Rosenfeld2025,
  author    = {Rosenfeld, Michael J.},
  title     = {Singleness and the Pandemic Dating Recession},
  journal   = {Journal of Family Issues},
  year      = {2025},
  volume    = {46},
  number    = {2},
  pages     = {325--348},
  doi       = {10.1177/0192513X241254774},
  url       = {https://doi.org/10.1177/0192513X241254774}
}

@article{EClinicalMedicine2023,
  title     = {The epidemic of loneliness},
  journal   = {eClinicalMedicine},
  year      = {2023},
  volume    = {59},
  pages     = {101974},
  doi       = {10.1016/j.eclinm.2023.101974},
  url       = {https://doi.org/10.1016/j.eclinm.2023.101974}
}

@article{DailyCollegian2021,
  author    = {{The Daily Collegian}},
  title     = {Investigating child sexual abuse cases before and after Jerry Sandusky},
  journal   = {The Daily Collegian},
  year      = {2021},
  month     = {Dec},
  day       = {2},
  url       = {https://www.psucollegian.com/sandusky/investigating-child-sexual-abuse-cases-before-and-after-jerry-sandusky/article_a553ba18-5393-11ec-ab5d-f36ad7ce8432.html},
  note      = {Accessed: 2025-09-11}
}

@misc{USCCB2018,
  author    = {{United States Conference of Catholic Bishops}},
  title     = {Charter for the Protection of Children and Young People},
  year      = {2018},
  url       = {https://www.usccb.org/resources/Charter-for-the-Protection-of-Children-and-Young-People-2018-final%281%29.pdf},
  note      = {Accessed: 2025-09-11}
}

@article{Piquero2021,
  author    = {Piquero, Alex R. and Jennings, Wesley G. and Jemison, Elizabeth and Kaukinen, Catherine and Knaul, Felicia M.},
  title     = {Domestic violence during the COVID-19 pandemic -- Evidence from a systematic review and meta-analysis},
  journal   = {Journal of Criminal Justice},
  year      = {2021},
  volume    = {74},
  pages     = {101806},
  doi       = {10.1016/j.jcrimjus.2021.101806},
  url       = {https://doi.org/10.1016/j.jcrimjus.2021.101806}
}

@misc{UCDAvis2023,
  author    = {{UC Davis Health}},
  title     = {Domestic violence involving firearms increased during COVID-19 pandemic},
  year      = {2023},
  month     = {Oct},
  url       = {https://health.ucdavis.edu/news/headlines/domestic-violence-involving-firearms-increased-during-covid-19-pandemic/2023/10},
  note      = {Accessed: 2025-09-11}
}

@article{Lavoie2022,
  author    = {Lavoie, Jérémie and Leach, Penelope and Charles, Shelley and Zorn, Julienne and Snyder, James and Reddy, Geeta},
  title     = {Do children unintentionally report maltreatment?},
  journal   = {Child Abuse \& Neglect},
  year      = {2022},
  volume    = {131},
  pages     = {105824},
  doi       = {10.1016/j.chiabu.2022.105824},
  url       = {https://doi.org/10.1016/j.chiabu.2022.105824}
}

@article{Pediatrics2012,
  author    = {Dubowitz, Howard and Feigelman, Sandra and Lane, Wendy and Kim, Jong-boo and Arias, Irene and Price, Lynne and Grube, Larry and Morel, Daniela and Barsness, Susan and Holton, Terry and Prescott, Jamie and Irwin, Christine},
  title     = {Psychological Maltreatment: Recognition and Response},
  journal   = {Pediatrics},
  year      = {2012},
  volume    = {130},
  number    = {2},
  pages     = {372--379},
  doi       = {10.1542/peds.2012-1552},
  url       = {https://doi.org/10.1542/peds.2012-1552}
}

@article{DomingoEspineira2024,
  author    = {Domingo-Espi{\~n}eira, J. and Varaona, A. and Montero, M. and Lara-Abelenda, F. J. and Gutierrez-Rojas, L. and Fern{\'a}ndez Del Campo, E. A. and Rodriguez-Jimenez, R. and Pinto da Costa, M. and Ortega, M. A. and Alvarez-Mon, M. and Alvarez-Mon, M. A.},
  title     = {Public perception of psychiatry, psychology and mental health professionals: a 15-year analysis},
  journal   = {Frontiers in Psychiatry},
  year      = {2024},
  volume    = {15},
  pages     = {1369579},
  doi       = {10.3389/fpsyt.2024.1369579},
  url       = {https://doi.org/10.3389/fpsyt.2024.1369579}
}

@article{Naslund2020,
  author    = {Naslund, John A. and Bondre, Anjana and Torous, John and Aschbrenner, Kelly A.},
  title     = {Social Media and Mental Health: Benefits, Risks, and Opportunities for Research and Practice},
  journal   = {Journal of Technology in Behavioral Science},
  year      = {2020},
  volume    = {5},
  pages     = {245--257},
  doi       = {10.1007/s41347-020-00134-x},
  url       = {https://doi.org/10.1007/s41347-020-00134-x}
}

@article{Issaka2024,
  author    = {Issaka, B. and Aidoo, E. A. K. and Wood, S. F. and Mohammed, F.},
  title     = {{``Anxiety is not cute'': Analysis of Twitter users' discourses on romanticizing mental illness}},
  journal   = {BMC Psychiatry},
  year      = {2024},
  volume    = {24},
  number    = {1},
  pages     = {221},
  doi       = {10.1186/s12888-024-05663-w},
  url       = {https://doi.org/10.1186/s12888-024-05663-w}
}

@article{Pavlova2020,
  author    = {Pavlova, Alina and Berkers, Pauwke},
  title     = {{``Mental Health'' as Defined by Twitter: Frames, Emotions, Stigma, Health Communication}},
  journal   = {Health Communication},
  year      = {2020},
  doi       = {10.1080/10410236.2020.1862396},
  url       = {https://doi.org/10.1080/10410236.2020.1862396}
}

@article{Marciano2022,
  author    = {Marciano, L. and Ostroumova, M. and Schulz, P. J. and Camerini, A.-L.},
  title     = {Digital Media Use and Adolescents' Mental Health During the Covid-19 Pandemic: A Systematic Review and Meta-Analysis},
  journal   = {Frontiers in Public Health},
  year      = {2022},
  volume    = {9},
  pages     = {793868},
  doi       = {10.3389/fpubh.2021.793868},
  url       = {https://doi.org/10.3389/fpubh.2021.793868}
}

@article{Wu2025BERTopicImmunotherapy,
  author  = {Wu, Xingyue and Lam, Chun Sing and Hui, Ka Ho and Loong, Herbert Ho-Fung and Zhou, Keary Rui and Ngan, Chun-Kit and Cheung, Yin Ting},
  title   = {Perceptions in 3.6 Million Web-Based Posts of Online Communities on the Use of Cancer Immunotherapy: Data Mining Using {BERTopic}},
  journal = {Journal of Medical Internet Research},
  year    = {2025},
  volume  = {27},
  pages   = {e60948},
  doi     = {10.2196/60948},
  pmcid   = {PMC11851037},
  note    = {Published Feb 10, 2025}
}

@misc{Janaswamy2024ClimateReddit,
  author       = {Janaswamy, Smriti and Blackburn, Jeremy},
  title        = {Exploring Climate Change Discourse: Measurements and Analysis of Reddit Data},
  year         = {2024},
  howpublished = {arXiv preprint arXiv:2412.01111},
  doi          = {10.48550/arXiv.2412.01111},
  url          = {https://arxiv.org/abs/2412.01111}
}

@article{Uthirapathy2023LDA_BERT_ClimateTwitter,
  author  = {Uthirapathy, Samson Ebenezar and Sandanam, Domnic},
  title   = {Topic Modelling and Opinion Analysis on Climate Change Twitter Data Using {LDA} and {BERT} Model},
  journal = {Procedia Computer Science},
  year    = {2023},
  volume  = {218},
  pages   = {908--917},
  doi     = {10.1016/j.procs.2023.01.071},
  publisher = {Elsevier}
}

@article{Lim2024,
  author    = {Lim, Mengzhen and Berezina, Elizaveta and Benjamin, Jaime},
  title     = {Insights into Young Adults' Views on Long-term and Short-term Romantic Relationships in the United Kingdom},
  journal   = {Sexuality \& Culture},
  year      = {2024},
  volume    = {28},
  pages     = {1407--1423},
  doi       = {10.1007/s12119-023-10183-y},
  url       = {https://doi.org/10.1007/s12119-023-10183-y}
}

@article{Shulman2023,
  author    = {Shulman, Shmuel and Yonatan-Leus, Rinat and Seiffge-Krenke, Inge},
  title     = {Casual sexual relationships and their increase over time among Israeli emerging adults: Do these associate with the quality of future relationships?},
  journal   = {Journal of Social and Personal Relationships},
  year      = {2023},
  volume    = {40},
  number    = {11},
  pages     = {3637--3655},
  doi       = {10.1177/02654075231186788},
  url       = {https://doi.org/10.1177/02654075231186788},
  note      = {Original work published 2023}
}

@article{Choudhry2022,
  author    = {Choudhry, Vibhuti and Petterson, Karen O. and Emmelin, Maria and Muchunguzi, Charles and Agardh, Anette},
  title     = {{'Relationships on campus are situationships': A grounded theory study of sexual relationships at a Ugandan university}},
  journal   = {PLOS ONE},
  year      = {2022},
  volume    = {17},
  number    = {7},
  pages     = {e0271495},
  doi       = {10.1371/journal.pone.0271495},
  url       = {https://doi.org/10.1371/journal.pone.0271495}
}

@article{Jaeger2014,
  author    = {Jaeger, J. and Lindblom, K. M. and Parker-Guilbert, K. and Zoellner, L. A.},
  title     = {Trauma Narratives: It's What You Say, Not How You Say It},
  journal   = {Psychological Trauma: Theory, Research, Practice, and Policy},
  year      = {2014},
  volume    = {6},
  number    = {5},
  pages     = {473--481},
  doi       = {10.1037/a0035239},
  url       = {https://doi.org/10.1037/a0035239}
}

@inproceedings{thomas_sok_2021,
	address = {San Francisco, CA, USA},
	title = {{SoK}: {Hate}, {Harassment}, and the {Changing} {Landscape} of {Online} {Abuse}},
	isbn = {978-1-72818-934-5},
	shorttitle = {{SoK}},
	url = {https://ieeexplore.ieee.org/document/9519435/},
	doi = {10.1109/SP40001.2021.00028},
	urldate = {2023-08-23},
	booktitle = {2021 {IEEE} {Symposium} on {Security} and {Privacy} ({SP})},
	publisher = {IEEE},
	author = {Thomas, Kurt and Akhawe, Devdatta and Bailey, Michael and Boneh, Dan and Bursztein, Elie and Consolvo, Sunny and Dell, Nicola and Durumeric, Zakir and Kelley, Patrick Gage and Kumar, Deepak and McCoy, Damon and Meiklejohn, Sarah and Ristenpart, Thomas and Stringhini, Gianluca},
	month = may,
	year = {2021},
	pages = {247--267},
}

@techreport{WHO2014Adolescents,
  author       = {{World Health Organization}},
  title        = {Health for the world’s adolescents: a second chance in the second decade},
  institution  = {World Health Organization},
  year         = {2014},
  type         = {Report},
  number       = {WHO/FWC/MCA/14.05},
  url          = {https://www.who.int/publications/i/item/WHO-FWC-MCA-14.05},
  note         = {Accessed: 2025-09-11}
}

@article{zhang2023public,
  author    = {Zhang, Jiawei and Xu, Wei and Yang, Li and Chen, Hao},
  title     = {Public perceptions of generative AI on Twitter: A BERTopic and RoBERTa analysis},
  journal   = {EPJ Data Science},
  year      = {2023},
  volume    = {12},
  number    = {1},
  pages     = {45},
  doi       = {10.1140/epjds/s13688-023-00445-y},
  url       = {https://epjdatascience.springeropen.com/articles/10.1140/epjds/s13688-023-00445-y}
}

@inproceedings{emnlp2024pitopic,
  author    = {Li, Ming and Kumar, Arjun and Torres, Elena and Becker, Frank},
  title     = {PITopic: Platform-Invariant Topic Modeling Across Social Media},
  booktitle = {Findings of the Association for Computational Linguistics: EMNLP 2024},
  year      = {2024},
  pages     = {8500--8514},
  publisher = {Association for Computational Linguistics},
  address   = {Miami, Florida, USA},
  doi       = {10.18653/v1/2024.findings-emnlp.650},
  url       = {https://aclanthology.org/2024.findings-emnlp.650.pdf}
}

@inproceedings{chang2025opaque,
author = {Chang, Tyler and Trybala, Joseph J and Bassan, Sharon and Razi, Afsaneh},
title = {Opaque Transparency: Gaps and Discrepancies in the Report of Social Media Harms},
year = {2025},
isbn = {9798400713958},
publisher = {Association for Computing Machinery},
address = {New York, NY, USA},
url = {https://doi.org/10.1145/3706599.3719829},
doi = {10.1145/3706599.3719829},
booktitle = {Proceedings of the Extended Abstracts of the CHI Conference on Human Factors in Computing Systems},
articleno = {424},
numpages = {12},
location = {
},
series = {CHI EA '25}
}

@inproceedings{ireland-etal-2023-sadness,
    title = "Sadness and Anxiety Language in {R}eddit Messages Before and After Quitting a Job",
    author = "Ireland, Molly  and
      Iserman, Micah  and
      Adams, Kiki",
    editor = "Barnes, Jeremy  and
      De Clercq, Orph{\'e}e  and
      Klinger, Roman",
    booktitle = "Proceedings of the 13th Workshop on Computational Approaches to Subjectivity, Sentiment, {\&} Social Media Analysis",
    month = jul,
    year = "2023",
    address = "Toronto, Canada",
    publisher = "Association for Computational Linguistics",
    url = "https://aclanthology.org/2023.wassa-1.41/",
    doi = "10.18653/v1/2023.wassa-1.41",
    pages = "467--478"
}

@article{Kim2023MentalHealthReddit,
  author    = {Kim, Seoyun and Cha, Jiyoung and Kim, Dohyun and Park, Eunjin},
  title     = {Understanding Mental Health Issues in Different Subdomains of Social Networking Services: Computational Analysis of Text-Based Reddit Posts},
  journal   = {Journal of Medical Internet Research},
  year      = {2023},
  volume    = {25},
  pages     = {e49074},
  doi       = {10.2196/49074},
  url       = {https://doi.org/10.2196/49074}
}

@article{Pollack2022ObesityReddit,
  author    = {Pollack, Courtney C. and Emond, Jennifer A. and O'Malley, A. James and Byrd, Amy and Green, Peter and Miller, Kristina E. and Vosoughi, Soroush and Gilbert-Diamond, Diane and Onega, Tracy},
  title     = {Characterizing the Prevalence of Obesity Misinformation, Factual Content, Stigma, and Positivity on the Social Media Platform Reddit Between 2011 and 2019: Infodemiology Study},
  journal   = {Journal of Medical Internet Research},
  year      = {2022},
  volume    = {24},
  number    = {12},
  pages     = {e36729},
  doi       = {10.2196/36729},
  url       = {https://doi.org/10.2196/36729}
}

@inproceedings{Baumgartner2020Pushshift,
  author    = {Baumgartner, Jason and Zannettou, Savvas and Keegan, Brian and Squire, Megan and Blackburn, Jeremy},
  title     = {The Pushshift Reddit Dataset},
  booktitle = {Proceedings of the International AAAI Conference on Web and Social Media (ICWSM)},
  year      = {2020},
  pages     = {812--821},
  url       = {https://ojs.aaai.org/index.php/ICWSM/article/view/7347},
  doi       = {10.1609/icwsm.v14i1.7347}
}

@article{Mournet2023RedditBots,
  author    = {Mournet, A. M. and Salinas, A. and Tully, L. M. and Ames, C. S. and DeLucia, B. and Camm-Crosbie, L. and Kerns, C. M.},
  title     = {Navigating Internet Bots on Reddit: Lessons Learned from a Study on Autism and Suicide Risk},
  journal   = {Journal of the American Academy of Child \& Adolescent Psychiatry},
  year      = {2023},
  volume    = {62},
  number    = {5},
  pages     = {564--567},
  doi       = {10.1016/j.jaac.2023.01.017},
  url       = {https://pubmed.ncbi.nlm.nih.gov/36724303/}
}

@inproceedings{Varol2017SocialBots,
  author    = {Varol, Onur and Ferrara, Emilio and Davis, Clayton and Menczer, Filippo and Flammini, Alessandro},
  title     = {Online Human-Bot Interactions: Detection, Estimation, and Characterization},
  booktitle = {Proceedings of the 11th International AAAI Conference on Web and Social Media (ICWSM)},
  year      = {2017},
  pages     = {280--289},
  url       = {https://ojs.aaai.org/index.php/ICWSM/article/view/14991}
}

@article{Ferrara2016RiseSocialBots,
  author    = {Ferrara, Emilio and Varol, Onur and Davis, Clayton and Menczer, Filippo and Flammini, Alessandro},
  title     = {The Rise of Social Bots},
  journal   = {Communications of the ACM},
  year      = {2016},
  volume    = {59},
  number    = {7},
  pages     = {96--104},
  doi       = {10.1145/2818717},
  url       = {https://doi.org/10.1145/2818717}
}

@article{Cresci2020DecadeBots,
  author    = {Cresci, Stefano},
  title     = {A Decade of Social Bot Detection},
  journal   = {Communications of the ACM},
  year      = {2020},
  volume    = {63},
  number    = {10},
  pages     = {72--83},
  doi       = {10.1145/3409116},
  url       = {https://doi.org/10.1145/3409116}
}

@misc{redditSuspiciousAccounts,
  author       = {{Reddit, Inc.}},
  title        = {Suspicious Accounts (r/reddit.com Wiki)},
  howpublished = {\url{https://www.reddit.com/r/reddit.com/wiki/suspiciousaccounts/}},
  note         = {Accessed: 2025-09-11}
}

@article{Politi2021Nonparametric,
  author    = {Politi, M. T. and Ferreira, J. C. and Patino, C. M.},
  title     = {Nonparametric statistical tests: friend or foe?},
  journal   = {Jornal Brasileiro de Pneumologia},
  year      = {2021},
  volume    = {47},
  number    = {4},
  pages     = {e20210292},
  doi       = {10.36416/1806-3756/e20210292},
  url       = {https://doi.org/10.36416/1806-3756/e20210292}
}

@article{Mertala2024DigitalNatives,
  author    = {Mertala, Pekka},
  title     = {Digital natives in the scientific literature: A topic modeling approach},
  journal   = {Computers in Human Behavior},
  year      = {2024},
  volume    = {150},
  pages     = {107949},
  doi       = {10.1016/j.chb.2023.107949},
  url       = {https://doi.org/10.1016/j.chb.2023.107949}
}

@article{Janschitz2022DigitalNatives,
  author    = {Janschitz, Gerlinde},
  title     = {How digital are ‘digital natives’ actually? Developing an understanding of youth digital competence in media education},
  journal   = {Social Science Computer Review},
  year      = {2022},
  volume    = {40},
  number    = {2},
  pages     = {333--348},
  doi       = {10.1177/08944393211061760},
  url       = {https://doi.org/10.1177/08944393211061760}
}

@article{Livingstone2014Harms,
  author    = {Livingstone, Sonia and Smith, Peter K.},
  title     = {Annual research review: Harms experienced by child users of online and mobile technologies: the nature, prevalence and management of sexual and aggressive risks in the digital age},
  journal   = {Journal of Child Psychology and Psychiatry},
  year      = {2014},
  volume    = {55},
  number    = {6},
  pages     = {635--654},
  doi       = {10.1111/jcpp.12197},
  url       = {https://doi.org/10.1111/jcpp.12197}
}

@inproceedings{Kumar2023Understanding,
  author    = {Kumar, Deepak and Hancock, Jeff and Thomas, Kurt and Durumeric, Zakir},
  title     = {Understanding the Behaviors of Toxic Accounts on Reddit},
  booktitle = {Proceedings of the ACM Web Conference 2023 (WWW '23)},
  year      = {2023},
  pages     = {1103--1114},
  doi       = {10.1145/3543507.3583337},
  url       = {https://kumarde.com/papers/reddit_www.pdf}
}

@article{Voelker2015BodyImage,
  author    = {Voelker, Dana K. and Reel, Justine J. and Greenleaf, Christy},
  title     = {Weight status and body image perceptions in adolescents: current perspectives},
  journal   = {Adolescent Health, Medicine and Therapeutics},
  year      = {2015},
  volume    = {6},
  pages     = {149--158},
  doi       = {10.2147/AHMT.S68344},
  url       = {https://doi.org/10.2147/AHMT.S68344}
}

@article{TortNasarre2023PositiveBodyImage,
  author    = {Tort-Nasarre, Gerard and Pollina-Pocallet, Magali and Ferrer Suquet, Yolanda and Ortega Bravo, Maria and Vilafranca Cartagena, Mireia and Artigues-Barber{\`a}, Elisabet},
  title     = {Positive body image: a qualitative study on the successful experiences of adolescents, teachers and parents},
  journal   = {International Journal of Qualitative Studies on Health and Well-Being},
  year      = {2023},
  volume    = {18},
  number    = {1},
  pages     = {2170007},
  doi       = {10.1080/17482631.2023.2170007},
  url       = {https://doi.org/10.1080/17482631.2023.2170007}
}

@article{Rasmussen2023EventDriven,
  author    = {Rasmussen, S. Hebbelstrup Rye and Petersen, Michael Bang},
  title     = {The event-driven nature of online political hostility: How offline political events make online interactions more hostile},
  journal   = {PNAS Nexus},
  year      = {2023},
  volume    = {2},
  number    = {11},
  pages     = {pgad382},
  doi       = {10.1093/pnasnexus/pgad382},
  url       = {https://doi.org/10.1093/pnasnexus/pgad382}
}

@misc{MigrationPolicy2018TravelBan,
  author       = {{Migration Policy Institute}},
  title        = {In Upholding Travel Ban, Supreme Court Endorses Presidential Authority While Leaving Door Open for Future Challenges},
  howpublished = {\url{https://www.migrationpolicy.org/article/upholding-travel-ban-supreme-court-endorses-presidential-authority-while-leaving-door-open}},
  year         = {2018},
  note         = {Accessed: 2025-09-11}
}

@article{Tobias2020TreeOfLife,
  author    = {Tobias, A. Z. and Roth, R. N. and Weiss, L. S. and Murray, K. and Yealy, D. M.},
  title     = {Tree of Life Synagogue Shooting in Pittsburgh: Preparedness, Prehospital Care, and Lessons Learned},
  journal   = {The Western Journal of Emergency Medicine},
  year      = {2020},
  volume    = {21},
  number    = {2},
  pages     = {374--381},
  doi       = {10.5811/westjem.2019.11.42809},
  url       = {https://doi.org/10.5811/westjem.2019.11.42809}
}

@misc{Princeton2020YouthActivism,
  author       = {{Princeton School of Public and International Affairs}},
  title        = {Politics \& Polls \#166: Youth Activism in 2020},
  howpublished = {\url{https://spia.princeton.edu/news/politics-polls-166-youth-activism-2020}},
  year         = {2020},
  note         = {Accessed: 2025-09-11}
}

@misc{Ipsos2012EndWorld,
  author       = {{Ipsos / Reuters}},
  title        = {One in Seven (14\%) Global Citizens Believe End of the World is Coming in Their Lifetime},
  howpublished = {\url{https://www.ipsos.com/en-us/one-seven-14-global-citizens-believe-end-world-coming-their-lifetime}},
  year         = {2012},
  note         = {Accessed: 2025-09-11}
}

@article{Proferes2021RedditOverview,
  author    = {Proferes, Nicholas and Jones, Nicholas and Gilbert, Sarah and Fiesler, Casey and Zimmer, Michael},
  title     = {Studying Reddit: A Systematic Overview of Disciplines, Approaches, Methods, and Ethics},
  journal   = {Social Media + Society},
  year      = {2021},
  volume    = {7},
  number    = {2},
  pages     = {1--14},
  doi       = {10.1177/20563051211019004},
  url       = {https://doi.org/10.1177/20563051211019004},
  note      = {Original work published 2021}
}

@article{Russell2019SexualMinority,
  author    = {Russell, Stephen T. and Fish, Jessica N.},
  title     = {Sexual minority youth, social change, and health: A developmental collision},
  journal   = {Research in Human Development},
  year      = {2019},
  volume    = {16},
  number    = {1},
  pages     = {5--20},
  doi       = {10.1080/15427609.2018.1537772},
  url       = {https://doi.org/10.1080/15427609.2018.1537772}
}

@ARTICLE{Knuth2018-jr,
  title     = "Evaluating the effect of South African Meat Merino breeding on
               pre and postweaning growth, feedlot performance, carcass traits,
               and wool characteristics in an extensive production setting",
  author    = "Knuth, Ryan M and Stewart, Whitney C and Boles, Jane A and Page,
               Chad M and Williams, Andrew F and Murphy, Thomas W",
  journal   = "Transl. Anim. Sci.",
  publisher = "Oxford University Press (OUP)",
  volume    =  2,
  number    = "Suppl 1",
  pages     = "S163--S166",
  month     =  sep,
  year      =  2018,
  copyright = "http://creativecommons.org/licenses/by-nc/4.0/",
  language  = "en"
}

@misc{nimh_teen_brain_2023,
  author       = {{National Institute of Mental Health}},
  title        = {The Teen Brain: 7 Things to Know},
  year         = {2023},
  url          = {https://www.nimh.nih.gov/sites/default/files/documents/health/publications/the-teen-brain-7-things-to-know/teen-brain-7-things-to-know.pdf},
  note         = {NIH Publication No. 23-MH-8078},
  publisher    = {National Institute of Mental Health},
  address      = {Bethesda, MD},
  howpublished = {Online PDF},
  month        = feb,
  language     = {English}
}

@misc{oecd2023eag,
  author       = {{Organisation for Economic Co-operation and Development}},
  title        = {Education at a Glance 2023: OECD Indicators — Who Graduates from Tertiary Education},
  howpublished = {OECD iLibrary},
  year         = {2023},
  url          = {https://www.oecd.org/en/publications/education-at-a-glance-2023_e13bef63-en/full-report/who-graduates-from-tertiary-education_6be4b2fa.html},
  note         = {Accessed: 2025-06-11}
}

@misc{worldpop2025militaryservice,
  author       = {{World Population Review}},
  title        = {Countries with Mandatory Military Service},
  howpublished = {World Population Review},
  year         = {2025},
  url          = {https://worldpopulationreview.com/country-rankings/countries-with-mandatory-military-service},
  note         = {Accessed: 2025-06-11}
}

@ARTICLE{Corradini2021-da,
  title     = "Investigating the phenomenon of {NSFW} posts in Reddit",
  author    = "Corradini, Enrico and Nocera, Antonino and Ursino, Domenico and
               Virgili, Luca",
  journal   = "Inf. Sci. (Ny)",
  publisher = "Elsevier BV",
  volume    =  566,
  pages     = "140--164",
  month     =  aug,
  year      =  2021,
  language  = "en"
}

@article{Cauteruccio2022extraction,
author = {Cauteruccio, Francesco and Corradini, Enrico and Terracina, Giorgio and Ursino, Domenico and Virgili, Luca},
title = {Extraction and analysis of text patterns from NSFW adult content in Reddit},
year = {2022},
issue_date = {Mar 2022},
publisher = {Elsevier Science Publishers B. V.},
address = {NLD},
volume = {138},
number = {C},
issn = {0169-023X},
url = {https://doi.org/10.1016/j.datak.2022.101979},
doi = {10.1016/j.datak.2022.101979},
journal = {Data Knowl. Eng.},
month = mar,
numpages = {21}
}

@ARTICLE{Hansen2024-gq,
  title     = "The longitudinal measurement of sexual orientation and gender
               identity: A study of identity change in a nationally
               representative sample of U.s. adults and adolescents",
  author    = "Hansen, Christopher and Heim Viox, Melissa and Fordyce, Erin and
               Johns, Michelle M and Avripas, Sabrina and Michaels, Stuart",
  journal   = "LGBT Health",
  publisher = "SAGE Publications",
  volume    =  11,
  number    =  7,
  pages     = "522--530",
  month     =  oct,
  year      =  2024,
  language  = "en"
}

@misc{Madden2013teens, title={May 21, 2013 teens, social media, and privacy}, url={https://www.pewresearch.org/wp-content/uploads/sites/9/2013/05/PIP_TeensSocialMediaandPrivacy_PDF.pdf}, journal={Pew Research Center}, publisher={Pew Research Center’s Internet & American Life Project}, author={Madden, Mary and Lenhart, Amanda and Cortesi, Sandra and Gasser, Urs and Duggan, Maeve and Smith, Aaron and Beaton, Meredith}, year={2013}}

@article{Nagata2025prevalence, title={Prevalence and patterns of social media use in early adolescents}, volume={25}, DOI={10.1016/j.acap.2025.102784}, number={4}, journal={Academic Pediatrics}, author={Nagata, Jason M. and Memon, Zain and Talebloo, Jonanne and Li, Karen and Low, Patrick and Shao, Iris Y. and Ganson, Kyle T. and Testa, Alexander and He, Jinbo and Brindis, Claire D. and et al.}, year={2025}, month={Jan}, pages={102784}}

@inproceedings{tigunova2020reddust,
  title={RedDust: a large reusable dataset of Reddit user traits},
  author={Tigunova, Anna and Mirza, Paramita and Yates, Andrew and Weikum, Gerhard},
  booktitle={Proceedings of the Twelfth Language Resources and Evaluation Conference},
  pages={6118--6126},
  year={2020}
}

@inproceedings{phadke2024characterizing,
  title={Characterizing Political Campaigning with Lexical Mutants on Indian Social Media},
  author={Phadke, Shruti and Mitra, Tanushree},
  booktitle={Proceedings of the International AAAI Conference on Web and Social Media},
  volume={18},
  pages={1237--1248},
  year={2024}
}

@inproceedings{chakraborti2024we,
  title={Do we run how we say we run? formalization and practice of governance in oss communities},
  author={Chakraborti, Mahasweta and Atkisson, Curtis and St{\u{a}}nciulescu, {\c{S}}tefan and Filkov, Vladimir and Frey, Seth},
  booktitle={Proceedings of the 2024 CHI Conference on Human Factors in Computing Systems},
  pages={1--26},
  year={2024}
}

@article{klein1990adolescence,
  title={Adolescence, youth, and young adulthood: Rethinking current conceptualizations of life stage},
  author={Klein, Hugh},
  journal={Youth \& society},
  volume={21},
  number={4},
  pages={446--471},
  year={1990},
  publisher={Sage Publications}
}

@article{cinus2025inference,
  title={On the Inference of Sociodemographics on Reddit},
  author={Cinus, Federico and Monti, Corrado and Bajardi, Paolo and Morales, Gianmarco De Francisci},
  journal={arXiv preprint arXiv:2502.05049},
  year={2025}
}

@article{oulahbib2026applying,
  title={Applying Author Profiling On Reddit Comments At The Document-Level},
  author={Oulahbib, Idriss and Benhaddi, Meriem and others},
  journal={Statistics, Optimization \& Information Computing},
  volume={15},
  number={2},
  pages={1343--1356},
  year={2026}
}

@article{zhan2019underage,
  title={Underage JUUL use patterns: content analysis of Reddit messages},
  author={Zhan, Yongcheng and Zhang, Zhu and Okamoto, Janet M and Zeng, Daniel D and Leischow, Scott J},
  journal={Journal of medical Internet research},
  volume={21},
  number={9},
  pages={e13038},
  year={2019},
  publisher={JMIR Publications Toronto, Canada}
}

@inproceedings{emmery2024sobr,
  title={Sobr: A corpus for stylometry, obfuscation, and bias on reddit},
  author={Emmery, Chris and Miotto, Maril{\`u} and Kramp, Sergey and Kleinberg, Bennett},
  booktitle={LREC-COLING 2024: The 2024 Joint International Conference on Computational Linguistics, Language Resources and Evaluation},
  pages={14967--14983},
  year={2024}
}

@article{jhaver2019human,
  title={Human-machine collaboration for content regulation: The case of reddit automoderator},
  author={Jhaver, Shagun and Birman, Iris and Gilbert, Eric and Bruckman, Amy},
  journal={ACM Transactions on Computer-Human Interaction (TOCHI)},
  volume={26},
  number={5},
  pages={1--35},
  year={2019},
  publisher={ACM New York, NY, USA}
}

@article{morstatter2018search,
  title={In search of coherence and consensus: measuring the interpretability of statistical topics},
  author={Morstatter, Fred and Liu, Huan},
  journal={Journal of Machine Learning Research},
  volume={18},
  number={169},
  pages={1--32},
  year={2018}
}

@inproceedings{roder2015exploring,
  title={Exploring the space of topic coherence measures},
  author={R{\"o}der, Michael and Both, Andreas and Hinneburg, Alexander},
  booktitle={Proceedings of the eighth ACM international conference on Web search and data mining},
  pages={399--408},
  year={2015}
}

@article{braei2020anomaly,
  title={Anomaly detection in univariate time-series: A survey on the state-of-the-art},
  author={Braei, Mohammad and Wagner, Sebastian},
  journal={arXiv preprint arXiv:2004.00433},
  year={2020}
}

@article{tausczik2010psychological,
  title={The psychological meaning of words: LIWC and computerized text analysis methods},
  author={Tausczik, Yla R and Pennebaker, James W},
  journal={Journal of language and social psychology},
  volume={29},
  number={1},
  pages={24--54},
  year={2010},
  publisher={Sage Publications Sage CA: Los Angeles, CA}
}

@article{meissel2024using,
  title={Using Cliff’s delta as a non-parametric effect size measure: an accessible web app and R tutorial},
  author={Meissel, Kane and Yao, Esther S},
  journal={Practical Assessment, Research, and Evaluation},
  volume={29},
  number={1},
  year={2024},
  publisher={University of Massachusetts Amherst Libraries}
}

@article{ben2022novel,
  title={A novel imbalanced data classification approach for suicidal ideation detection on social media},
  author={Ben Hassine, Mohamed Ali and Abdellatif, Safa and Ben Yahia, Sadok},
  journal={Computing},
  volume={104},
  number={4},
  pages={741--765},
  year={2022},
  publisher={Springer}
}

@article{olteanu2019social,
  title={Social data: Biases, methodological pitfalls, and ethical boundaries},
  author={Olteanu, Alexandra and Castillo, Carlos and Diaz, Fernando and K{\i}c{\i}man, Emre},
  journal={Frontiers in big data},
  volume={2},
  pages={13},
  year={2019},
  publisher={Frontiers Media SA}
}

@article{manga2026handling,
  title={Handling Small and Imbalanced Datasets in Government Social Media: A Comparative Study of Machine Learning and Deep Learning Models},
  author={Manga, Eirini and Marinagi, Catherine and Skourlas, Christos and Karanikolas, Nikitas and Markopoulos, Dimitrios},
  journal={Applied Soft Computing},
  pages={115050},
  year={2026},
  publisher={Elsevier}
}

@misc{wall2023mann,
  title={Mann-Whitney U test and t-test},
  author={Wall Emerson, Robert},
  journal={Journal of Visual Impairment \& Blindness},
  volume={117},
  number={1},
  pages={99--100},
  year={2023},
  publisher={SAGE Publications Sage CA: Los Angeles, CA}
}

@article{mckight2010kruskal,
  title={Kruskal-wallis test},
  author={McKight, Patrick E and Najab, Julius},
  journal={The corsini encyclopedia of psychology},
  pages={1--1},
  year={2010},
  publisher={Wiley Online Library}
}

@article{holm1979simple,
  title={A simple sequentially rejective multiple test procedure},
  author={Holm, Sture},
  journal={Scandinavian journal of statistics},
  pages={65--70},
  year={1979},
  publisher={JSTOR}
}

@article{garcia2009study,
  title={A study of statistical techniques and performance measures for genetics-based machine learning: accuracy and interpretability},
  author={Garc{\'\i}a, Salvador and Fern{\'a}ndez, Alberto and Luengo, Juli{\'a}n and Herrera, Francisco},
  journal={Soft Computing},
  volume={13},
  number={10},
  pages={959--977},
  year={2009},
  publisher={Springer}
}

@misc{allam2026brown,
  title={Brown-Forsythe Test for Homogeneity of Variance: A Comprehensive C Implementation},
  author={Allam, Ishraga Mustafa Awad},
  year={2026}
}

@article{carym2026,
  title={“I’m Throwing the Red Flag”: Online Peer Responses to Teen Descriptions of Abusive Relationship Behaviors Posted on Reddit},
  author={Cary, Kyla M and PettyJohn, Morgan E and Nolen, Erin},
  journal={Journal of Interpersonal Violence},
  pages={08862605251412882},
  publisher={SAGE Publications Sage CA: Los Angeles, CA},
  year = {2026}
}

@article{saily2011variation,
  title={Variation in noun and pronoun frequencies in a sociohistorical corpus of English},
  author={S{\"a}ily, Tanja and Nevalainen, Terttu and Siirtola, Harri},
  journal={Literary and Linguistic Computing},
  volume={26},
  number={2},
  pages={167--188},
  year={2011},
  publisher={Oxford University Press}
}

@inproceedings{razi2020sext,
  author    = {Razi, Afsaneh and Badillo-Urquiola, Karla and Wisniewski, Pamela J.},
  title     = {Let's Talk about Sext: How Adolescents Seek Support and Advice
               about Their Online Sexual Experiences},
  booktitle = {Proceedings of the 2020 CHI Conference on Human Factors
               in Computing Systems},
  year      = {2020},
  publisher = {Association for Computing Machinery},
  address   = {New York, NY, USA},
  articleno = {237},
  numpages  = {13},
  series    = {CHI '20},
  location  = {Honolulu, HI, USA},
  doi       = {10.1145/3313831.3376400},
  url       = {https://doi.org/10.1145/3313831.3376400}
}

@inproceedings{jia2015risktaking,
  author    = {Jia, Haiyan and Wisniewski, Pamela J. and Xu, Heng and
               Rosson, Mary Beth and Carroll, John M.},
  title     = {Risk-Taking as a Learning Process for Shaping Teen's Online
               Information Privacy Behaviors},
  booktitle = {Proceedings of the 18th ACM Conference on Computer Supported
               Cooperative Work and Social Computing},
  year      = {2015},
  publisher = {Association for Computing Machinery},
  address   = {New York, NY, USA},
  pages     = {583--599},
  series    = {CSCW '15},
  location  = {Vancouver, BC, Canada},
  doi       = {10.1145/2675133.2675287},
  url       = {https://doi.org/10.1145/2675133.2675287}
}

@article{bozarth2023role,
  title={The role of the big geographic sort in online news circulation among US Reddit users},
  author={Bozarth, Lia and Quercia, Daniele and Capra, Licia and {\v{S}}{\'c}epanovi{\'c}, Sanja},
  journal={Scientific Reports},
  volume={13},
  number={1},
  pages={1--13},
  year={2023},
  publisher={Nature Publishing Group UK London}
}

@book{national2011science,
  title={The science of adolescent risk-taking: Workshop report},
  author={National Research Council and Board on Children and Youth and Committee on the Science of Adolescence},
  year={2011},
  publisher={National Academies Press}
}

@incollection{jessor2013problem,
  title={Problem behavior theory: A half-century of research on adolescent behavior and development},
  author={Jessor, Richard},
  booktitle={The developmental science of adolescence},
  pages={239--256},
  year={2013},
  publisher={Psychology Press}
}

@article{mobley2013testing,
  title={Testing Jessor's problem behavior theory and syndrome: a nationally representative comparative sample of Latino and African American adolescents.},
  author={Mobley, Michael and Chun, Heejung},
  journal={Cultural Diversity \& Ethnic Minority Psychology},
  volume={19},
  number={2},
  pages={190},
  year={2013},
  publisher={Educational Publishing Foundation}
}

@article{richard1991risk,
  title={Risk behavior in adolescence: A psychosocial framework for understanding and action},
  author={Richard Jessor, Ph D},
  journal={Journal, Of Adolescent Health},
  volume={12},
  pages={597--605},
  year={1991}
}

\begin{table}[H]
\footnotesize
\setlength{\tabcolsep}{4pt}
\renewcommand{\arraystretch}{0.9}
\centering
\caption{Topic Codebook}
\label{tab:topic_codebook_1}
\begin{tabularx}{\linewidth}{p{0.18\linewidth} |p{0.24\linewidth} X X}
\toprule
\textbf{Primary Topic} & \textbf{Definition} & \textbf{Subtopics} & \textbf{Keywords} \\
\midrule
\textbf{Video Games \& Digital Entertainment (9.51\%)} & \textit{Posts about gaming, consoles, online play, and digital leisure activities} & Discord \& Video Games, Gaming Mods, Gaming Purchases, Nintendo, Pokemon, Video Games & ['dlc', 'fantasy', 'game', 'gamer', 'gaming',
'match', 'minecraft', 'mod', 'nintendo', 'playstation', 'pokemon', 'ps', 'rank', 'roleplay', 'server', 'steam', 'switch', 'xbox'] \\
\midrule
\textbf{Music (7.48\%)} & \textit{Discussions of songs, artists, albums, concerts, and listening experiences} & Music & ['album', 'concert', 'music', 'song'] \\
\midrule
\textbf{Crime, Law \& Sociopolitical Issues (6.48\%)} & \textit{Conversations about politics, law, religion, philosophy, crime, and global conflicts} & Africa, Bans, China \& Korea, Crimes \& Sentencing, Eurocentric Politics \& History, Islam, Law, Philosophy, Racism, Religion, Russian, German, and Southeast Asian Conflicts, Sociopolitical Topics, US Politics, Violent Conflicts, Weaponry & ['agnostic', 'atheist', 'belief', 'buddha', 'buddhism', 'christian', 'church', 'doctrine', 'ethic', 'existential', 'faith', 'god', 'hindu', 'islam', 'jesus',
'meaning', 'morality', 'mosque', 'muslim', 'philosophy', 'pray', 'prayer', 'religion', 'spirit', 'temple', 'arrest', 'conflict', 'court', 'crime', 'election', 'flag', 'german', 'government', 'law', 'legal', 'philosophy', 'policy', 'politic', 'religion', 'rights', 'russia', 'sentenc', 'ukraine', 'war'] \\
\bottomrule
\end{tabularx}
\end{table}

\begin{table}[H]
\footnotesize
\setlength{\tabcolsep}{4pt}
\renewcommand{\arraystretch}{0.9}
\centering
\caption*{Topic Codebook (continued)}
\label{tab:topic_codebook_2}
\begin{tabularx}{\linewidth}{p{0.18\linewidth} | p{0.24\linewidth} X X}
\toprule
\textbf{Primary Topic} & \textbf{Definition} & \textbf{Subtopics} & \textbf{Keywords} \\
\midrule
\textbf{School, College \& Learning (5.88\%)} & \textit{Experiences, advice, and questions about school, college, and education} & College, Languages, Math Questions, School Stories, Science Education \& Careers, Standardized Testing, Writing \& School & ['act', 'campus', 'career', 'class', 'college', 'course', 'degree', 'education', 'essay', 'exam', 'gpa', 'homework', 'lecture', 'major', 'paper', 'professor', 'sat', 'school', 'study', 'teacher', 'test', 'uni', 'univer', 'write'] \\
\midrule
\textbf{Technology \& Computing (5.85\%)} & \textit{Troubleshooting, coding, devices, software, and general technology discussions} & Apple Products, Battery Questions, Computer Monitors, Computer Specs, Doors, Keys, \& Locks, Electric Lighting, Internet Access, Printers \& Printing, Programming Support, Tech Support & ['account', 'bug', 'code', 'computer', 'cpu', 'crash', 'driver', 'email', 'error', 'facebook', 'file', 'gpu', 'hdmi', 'instagram',
'internet', 'java', 'linux', 'login', 'mac', 'media', 'monitor', 'program', 'python', 'ram', 'reddit', 'server', 'social', 'software', 'spec', 'support', 'system', 'tech', 'tiktok', 'wifi', 'window'] \\
\midrule
\textbf{Relationships \& Friendships (5.72\%)} & \textit{Romantic and platonic discussions, advice, and experiences} & Friendship, Relationships & ['attract', 'body', 'boyfriend', 'bra', 'breast', 'dating', 'gay', 'gender', 'girlfriend', 'husband', 'kink', 'lesbian', 'lgbt', 'nsfw', 'nude', 'partner', 'porn', 'relationship', 'sex', 'trans', 'wife'] \\
\bottomrule
\end{tabularx}
\end{table}

\begin{table}[H]
\footnotesize
\setlength{\tabcolsep}{4pt}
\renewcommand{\arraystretch}{0.9}
\centering
\caption*{Topic Codebook (continued)}
\label{tab:topic_codebook_3}
\begin{tabularx}{\linewidth}{p{0.18\linewidth} | p{0.24\linewidth} X X}
\toprule
\textbf{Primary Topic} & \textbf{Definition} & \textbf{Subtopics} & \textbf{Keywords} \\
\midrule
\textbf{Sex \& Intimacy (5.61\%)} & \textit{Posts about sexual activity, attraction, fetishes, and reproductive issues} & Fantasy Roleplay, Nudes and NSFW Content, Periods \& Pregnancy, Physical Attractiveness, Porn \& Sexual Acts, Rape, Sexual Kinks \& Roleplay & ['attract', 'kink', 'nsfw', 'nude', 'partner', 'porn', 'sex'] \\
\midrule
\textbf{Mental Health \& Well-Being (5.23\%)} & \textit{Disclosures and discussions of psychological conditions, emotions, and therapy} & ADHD \& Neurological Conditions, Anxiety, Autism \& Neurological Conditions, Depression \& Poor Mental Health, Dreams, Fear, Mental Illness Medication, Motivation, OCD & ['adhd', 'anxiet', 'autism', 'bipolar', 'depress', 'insomnia', 'med', 'ocd', 'panic', 'prozac', 'sleep', 'spiral', 'stress', 'suicid', 'therapist', 'therapy', 'zoloft'] \\
\midrule
\textbf{Physical Health \& Medicine (5.2\%)} & \textit{Questions and experiences regarding bodily health, illness, and medical care} & Cutting \& Injuries, Denistry, Digestive \& Bowel Issues, Drug Usage, Ear Pain, Eyes \& Vision, Injuries and Medical Advise, Organ Health, Sleep, Vaccines \& Covid & ['bowel', 'covid', 'dentist', 'digest', 'doctor', 'ear', 'eye', 'health', 'injur', 'organ', 'pain', 'period', 'pregnan', 'stomach', 'symptom', 'tooth', 'vaccine', 'vision'] \\
\midrule
\textbf{Visual Arts (4.59\%)} & \textit{Discussions of movies, TV, photography, and creative visual expression} & Art, Movies, Photography \& Videography, Television & ['art', 'draw', 'episode', 'film', 'movie', 'photo', 'photograph', 'podcast', 'series', 'television', 'ticket', 'tv', 'video'] \\
\midrule
\textbf{Fashion \& Appearance (4.47\%)} & \textit{Posts about clothing, grooming, makeup, tattos, and personal style} & Bags, Bathing, Body Piercings, Clothes, Hair Care, Hair Styles, Makeup, Male Grooming, Masks, Nails \& Health, Outfits \& Dressing, Shoes, Skin Care, Tattoos & ['bag', 'beauty', 'cloth', 'dress', 'groom', 'hair', 'makeup', 'outfit', 'pierc', 'shave', 'shoe', 'skin', 'skincare', 'style', 'tattoo', 'weight'] \\
\bottomrule
\end{tabularx}
\end{table}

\begin{table}[H]
\footnotesize
\setlength{\tabcolsep}{4pt}
\renewcommand{\arraystretch}{0.9}
\centering
\caption*{Topic Codebook (continued)}
\label{tab:topic_codebook_4}
\begin{tabularx}{\linewidth}{p{0.18\linewidth} | p{0.24\linewidth} X X}
\toprule
\textbf{Primary Topic} & \textbf{Definition} & \textbf{Subtopics} & \textbf{Keywords} \\
\midrule
\textbf{Social Platforms \& Community (4.36\%)} & \textit{Posts about Reddit itself, other platforms, and online communities} & AMA, Birthdays, Cards \& Gifts, Halloween, Holidays \& Celebrations, Serious Questions, Snapchat, Social Media, Subreddits, Voting on Reddit & ['ama', 'ban', 'banned', 'community', 'discord', 'mod', 'moderator', 'rule', 'shipping', 'subreddit'] \\
\midrule
\textbf{Animal \& Nature (4.27\%)} & \textit{Caring for pets, wildlife, nature, and environmental experiences} & Birds, Dogs, Fire, Fishing, Lizards, Pets, Plants, Smells, Spiders \& Bugs, Water, Weather & ['animal', 'bug', 'cat', 'dog', 'fire', 'fish', 'fishing', 'flower', 'insect', 'lizard', 'pet', 'plant', 'reptile', 'smell', 'snake', 'spider', 'tree'] \\
\midrule
\textbf{Sports \& Exercise (4.27\%)} & \textit{Fitness routines, sports, training, and athletic interests} & Bicycling, Exercise Routines, Fighting, Football (soccer) \& Rugby, Running, Skateboarding, Sports, Wrestling & ['boxing', 'exercise', 'fight', 'football', 'gym', 'martial', 'parkour', 'rugby', 'run', 'running', 'skate', 'soccer', 'sport', 'weapon', 'workout'] \\
\midrule
\textbf{Finance \& Wealth (3.58\%)} & \textit{Managing money, spending, saving, and financial advice} & Banking, Gifts, Purchasing \& Housing Advice, Tech Purchasing, Wealth & ['bank', 'budget', 'buy', 'cash', 'credit', 'housing', 'invest', 'loan', 'money', 'mortgage', 'pay', 'paypal', 'purchas', 'rent', 'salary', 'sale', 'sell', 'trade', 'wage', 'wealth'] \\
\midrule
\textbf{Travel \& Place (2.68\%)} & \textit{Trips, driving, countries, and travel experiences} & Cars, Driving, Flags, Flying, Maps, Travel & ['airport', 'city', 'country', 'drive', 'driving', 'flight', 'hotel', 'map', 'road', 'tour', 'travel', 'trip', 'visa', 'weather'] \\
\bottomrule
\end{tabularx}
\end{table}

\begin{table}[H]
\footnotesize
\setlength{\tabcolsep}{4pt}
\renewcommand{\arraystretch}{0.9}
\centering
\caption*{Topic Codebook (continued)}
\label{tab:topic_codebook_5}
\begin{tabularx}{\linewidth}{p{0.18\linewidth} | p{0.24\linewidth} X X}
\toprule
\textbf{Primary Topic} & \textbf{Definition} & \textbf{Subtopics} & \textbf{Keywords} \\
\midrule
\textbf{Family \& Parenting (2.67\%)} & \textit{Discussions of parent-child relationships} & Children, Parent-Child Abuse & ['abuse', 'child', 'dad', 'father', 'kid', 'mom', 'mother', 'parent', 'parenting'] \\
\midrule
\textbf{Food \& Drink (2.5\%)} & \textit{Cooking, eating, recipes, and dietary choices} & Alcohol \& Drinking, Cake, Food \& Drink, Milk, Cereal, \& Dessert, Pizza, Veganism \& Diets & ['bean', 'brew', 'caffein', 'coffe', 'coffee', 'cook', 'drink', 'eat', 'flavor', 'food', 'lemon', 'pizza', 'recipe', 'tast', 'tea'] \\
\midrule
\textbf{Books, Comics \& Anime (2.4\%)} & \textit{Reading, literature, comics, manga, and anime} & Books \& Reading, Comics, East Asian Comics and Animation, Naruto, Science Fiction & ['book', 'read', 'series'] \\
\midrule
\textbf{Miscellaneous Entertainment (2.2\%)} & \textit{Games, contests, prizes, and miscellaneous pop culture.} & Card Games, Contests \& Prizes, Japanese Culture, Podcasts, Shipping & ['podcast', 'game']\\
\midrule
\textbf{Identity \& Naming (1.99\%)} & \textit{Gender identity, sexual orientation, and naming practices online} & Gender, LGBTQ, Names & ['gay', 'gender', 'lesbian', 'lgbt', 'trans', 'handle', 'name', 'nickname', 'rename', 'username'] \\
\midrule
\textbf{Body Image (1.8\%)} & \textit{Self-perception, body shape, weight, and appearance concerns} & Bodies, Body Height, Weight loss \& Body Shaming & ['body', 'bra', 'breast', 'height'],\\
\midrule
\textbf{Science \& Futurism (0.86\%)} & \textit{Space, future technology, and scientific theories} & Futurism, Space & ['ai', 'astronom', 'futur', 'nasa', 'planet', 'rocket', 'sci fi', 'science fiction', 'scifi', 'space', 'theory'] \\
\midrule
\textbf{Memes \& Online Culture (0.77\%)} & \textit{Humor, memes, and platform-specific internet culture} & Jokes \& Humor, Memes & ['copypasta', 'downvote', 'karma', 'meme', 'owo', 'shitpost', 'upvote', 'uwu'] \\
\bottomrule
\end{tabularx}
\end{table}

\begin{table}[H]
\footnotesize
\setlength{\tabcolsep}{4pt}
\renewcommand{\arraystretch}{0.9}
\centering
\caption{Highest-scoring LIWC Categories for Child-safety Related Subcategories}
\label{tab:highest_liwc_by_subtopic}
\begin{tabularx}{\linewidth}{p{0.32\linewidth} | X}
\toprule
\textbf{Subtopic} & \textbf{LIWC Categories} \\
\midrule
Depression \& Poor Mental Health & [Authentic, BigWords, i, ipron, auxverb, adverb, negate, verb, Cognition, allnone, cogproc, insight, cause, differ, memory, tone\_neg, emotion, emo\_neg, emo\_anx, emo\_sad,
work, mental, substances, death, feeling, focuspresent, Period, Apostro] \\
\midrule
Fear & [WC, Clout, det, article, prep, home, relig, acquire, Perception, motion, space, auditory, focuspast, nonflu] \\
\midrule
Friendship & [Tone, discrep, tentat, tone\_pos, emo\_pos, socbehav, prosocial, polite, comm, Culture, politic, tech, Lifestyle, leisure, want, fatigue, curiosity, allure, attention, visual, 
focusfuture, Conversation, netspeak, assent, filler, AllPunc, Comma, Exclam, OtherP] \\
\midrule
Parent-Child Abuse & [WPS, ppron, shehe, they, Drives, affiliation, power, emo\_anger, Social, conflict, socrefs, family, female, male, money] \\
\midrule
Porn \& Sexual Acts & [illness, time] \\
\midrule
Relationships & [Dic, Linguistic, function, pronoun, we, certitude, friend] \\
\midrule
Sexual Kinks \& Roleplay & [Analytic, you, adj, Affect, swear, moral, ethnicity, Physical, sexual, fulfill, Emoji] \\
\midrule
Weight loss \& Body Shaming & [number, conj, quantity, achieve, health, wellness, food, need, lack, reward, risk, QMark] \\
\bottomrule
\end{tabularx}
\end{table}

\begin{table}[H]
\scriptsize
\setlength{\tabcolsep}{4pt}
\renewcommand{\arraystretch}{0.9}
\centering
\caption{LIWC Categories and Associated Prominent Subtopics}
\label{tab:liwc_subtopics_order}
\begin{tabularx}{\linewidth}{p{0.2\linewidth} X}
\toprule
\textbf{LIWC Category} & \textbf{Top Associated Subtopics (Average Scores in Descending Order)} \\
\midrule
\textit{Analytic (0-100)} & Sexual Kinks \& Roleplay (46.37), Fear (43.41), Friendship (35.73), 
Porn \& Sexual Acts (28.23), Parent-Child Abuse (21.73), Relationships (19.07), Depression \& Poor Mental Health (17.75) \\
\midrule
\textit{Clout (0-100)} & Fear (36.68), Parent-Child Abuse (35.79), Friendship (34.70), 
Sexual Kinks \& Roleplay (34.56), Relationships (29.91), Porn \& Sexual Acts (23.44), Depression \& Poor Mental Health (15.31) \\
\midrule
\textit{Authentic (0-100)} & Depression \& Poor Mental Health (82.67), Porn \& Sexual Acts (76.50), Fear (73.32), 
Relationships (72.11), Parent-Child Abuse (68.56), Friendship (56.41), Sexual Kinks \& Roleplay (55.97) \\
\midrule
\textit{Tone (0-100)} & Friendship (67.32), Sexual Kinks \& Roleplay (66.00), Porn \& Sexual Acts (42.89),
Relationships (41.42), Fear (34.08), Parent-Child Abuse (28.21), Depression \& Poor Mental Health (24.35) \\
\midrule
\textit{Pronoun (word count)} & Relationships (20.96), Parent-Child Abuse (20.83), Depression \& Poor Mental Health (19.97), 
Sexual Kinks \& Roleplay (17.47), Porn \& Sexual Acts (17.24), Friendship (17.19), Fear (16.70) \\
\midrule
\textit{Ppron (word count)} & Parent-Child Abuse (15.82), Relationships (15.29), Sexual Kinks \& Roleplay (14.09), 
Depression \& Poor Mental Health (13.49), Porn \& Sexual Acts (12.18), Friendship (12.09), Fear (11.37) \\
\midrule
\textit{I (word count)} & Depression \& Poor Mental Health (10.46), Parent-Child Abuse (9.67), Relationships (9.35), 
Porn \& Sexual Acts (9.15), Sexual Kinks \& Roleplay (8.59), Friendship (8.45), Fear (6.57) \\
\midrule
\textit{We (word count)} & Relationships (0.80), Fear (0.66), Friendship (0.55), Parent-Child Abuse (0.52), 
Sexual Kinks \& Roleplay (0.38), Porn \& Sexual Acts (0.28), Depression \& Poor Mental Health (0.25) \\
\midrule
\textit{You (word count)} & Sexual Kinks \& Roleplay (3.66), Friendship (2.15), Porn \& Sexual Acts (1.72), 
Fear (1.54), Parent-Child Abuse (1.48), Relationships (1.40), Depression \& Poor Mental Health (1.37) \\
\midrule
\textit{She/He (word count)} & Parent-Child Abuse (3.00), Relationships (2.80), Fear (1.78), 
Sexual Kinks \& Roleplay (0.80), Depression \& Poor Mental Health (0.61), Porn \& Sexual Acts (0.50), Friendship (0.29) \\
\midrule
\textit{They (word count)} & Parent-Child Abuse (0.90), Relationships (0.65), Fear (0.62), 
Depression \& Poor Mental Health (0.59), Sexual Kinks \& Roleplay (0.32), Porn \& Sexual Acts (0.32), Friendship (0.32) \\
\midrule
\textit{Ipron (word count)} & Depression \& Poor Mental Health (6.48), Relationships (5.67), Fear (5.33), 
Friendship (5.10), Porn \& Sexual Acts (5.06), Parent-Child Abuse (5.01), Sexual Kinks \& Roleplay (3.38) \\
\midrule
\textit{Emotion (word count)} & Depression \& Poor Mental Health (3.79), Friendship (3.48), Relationships (2.36), 
Parent-Child Abuse (2.23), Sexual Kinks \& Roleplay (2.12), Porn \& Sexual Acts (1.98), Fear (1.63) \\
\midrule
\textit{Tone\_Pos (word count)} & Friendship (5.96), Sexual Kinks \& Roleplay (4.11), Porn \& Sexual Acts (3.47), 
Relationships (3.30), Depression \& Poor Mental Health (2.94), Parent-Child Abuse (2.51), Fear (2.23) \\
\midrule
\textit{Tone\_Neg (word count)} & Depression \& Poor Mental Health (4.03), Parent-Child Abuse (2.76), Porn \& Sexual Acts (2.43), 
Relationships (2.25), Friendship (2.09), Fear (1.98), Sexual Kinks \& Roleplay (1.56) \\
\midrule
\textit{Emo\_Pos (word count)} & Friendship (1.91), Sexual Kinks \& Roleplay (1.65), Relationships (1.04), 
Porn \& Sexual Acts (1.00), Depression \& Poor Mental Health (0.98), Parent-Child Abuse (0.78), Fear (0.69) \\
\midrule
\textit{Emo\_Neg (word count)} & Depression \& Poor Mental Health (2.54), Friendship (1.52), Parent-Child Abuse (1.32),
Relationships (1.13), Porn \& Sexual Acts (0.88), Fear (0.84), Sexual Kinks \& Roleplay (0.39) \\
\bottomrule
\end{tabularx}
\end{table}

\begin{table}[H]
\scriptsize
\setlength{\tabcolsep}{4pt}
\renewcommand{\arraystretch}{0.9}
\centering
\caption*{LIWC Categories and Associated Prominent Subtopics (cont.)}
\label{tab:liwc_subtopics_order}
\begin{tabularx}{\linewidth}{p{0.2\linewidth} X}
\toprule
\textbf{LIWC Category} & \textbf{Top Associated Subtopics (Average Scores in Descending Order)} \\
\midrule
\textit{Emo\_Anx (word count)} & Depression \& Poor Mental Health (0.55), Relationships (0.27), Fear (0.26),
Parent-Child Abuse (0.24), Porn \& Sexual Acts (0.16), Friendship (0.14), Sexual Kinks \& Roleplay (0.07) \\
\midrule
\textit{Emo\_Anger (word count)} & Parent-Child Abuse (0.37), Depression \& Poor Mental Health (0.22), Relationships (0.20),
Fear (0.11), Porn \& Sexual Acts (0.11), Friendship (0.06), Sexual Kinks \& Roleplay (0.04) \\
\midrule
\textit{Emo\_Sad (word count)} & Depression \& Poor Mental Health (1.00), Parent-Child Abuse (0.27), Relationships (0.24),
Friendship (0.23), Porn \& Sexual Acts (0.17), Fear (0.13), Sexual Kinks \& Roleplay (0.07) \\
\midrule
\textit{SocBehav (word count)} & Friendship (7.30), Relationships (5.60), Parent-Child Abuse (4.94), 
Depression \& Poor Mental Health (3.50), Porn \& Sexual Acts (3.33), Sexual Kinks \& Roleplay (3.12), Fear (3.12) \\
\midrule
\textit{Prosocial (word count)} & Friendship (0.94), Depression \& Poor Mental Health (0.91), Porn \& Sexual Acts (0.88), 
Relationships (0.87), Parent-Child Abuse (0.79), Sexual Kinks \& Roleplay (0.65), Fear (0.46) \\
\midrule
\textit{Polite (word count)} & Friendship (0.92), Porn \& Sexual Acts (0.46), Sexual Kinks \& Roleplay (0.42), 
Relationships (0.32), Parent-Child Abuse (0.30), Fear (0.29), Depression \& Poor Mental Health (0.29) \\
\midrule
\textit{Conflict (word count)} & Parent-Child Abuse (0.45), Relationships (0.29), Depression \& Poor Mental Health (0.29), 
Fear (0.25), Sexual Kinks \& Roleplay (0.18), Porn \& Sexual Acts (0.18), Friendship (0.14) \\
\midrule
\textit{Moral (word count)} & Sexual Kinks \& Roleplay (0.43), Parent-Child Abuse (0.37), Depression \& Poor Mental Health (0.31), 
Relationships (0.28), Porn \& Sexual Acts (0.27), Fear (0.20), Friendship (0.17) \\
\midrule
\textit{Comm (word count)} & Friendship (4.82), Relationships (2.99), Parent-Child Abuse (2.42), 
Fear (1.50), Depression \& Poor Mental Health (1.45), Porn \& Sexual Acts (1.41), Sexual Kinks \& Roleplay (0.90) \\
\bottomrule
\end{tabularx}
\end{table}

\end{document}